\documentclass{article}
\usepackage{graphicx}
\usepackage[sort&compress, numbers]{natbib}
\usepackage{hyperref}
\usepackage{amsmath}
\usepackage{amssymb}
\usepackage{mathtools}
\usepackage{algorithm}
\usepackage{algpseudocode}
\usepackage{geometry}
\usepackage{longtable}
\usepackage{booktabs}
\usepackage{pgfplots}
\usepackage[pagewise]{lineno}
\numberwithin{algorithm}{section}
\usepackage{wrapfig}
\def\r{\mathbf{r}}

\def\e{\text{e}}
\def\d{\text{d}}

\pgfplotsset{compat=1.18}
\begin{document}

\begin{titlepage}
    \begin{center}
        \vspace*{0.2cm}
        
        \large
        \textbf{A global inverse-problem approach to\\ quantitative photo-switching optoacoustic mesoscopy}
        
        \vspace{1.5cm}
        
        \normalsize

		\centerline{\scshape Yan Liu}
		\medskip
		{\footnotesize
            \centerline{yan.liu@epfl.ch}
		      \centerline{Biomedical Imaging Group, École polytechnique fédérale de Lausanne}
		    \centerline{1015 Lausanne, Switzerland}
		} % Do not forget to end the {\footnotesize by the sign }
        
        \medskip
        
        \centerline{\scshape Jonathan Chuah}
		\medskip
        {\footnotesize
            \centerline{jonathan.chuahwenjie@epfl.ch}
		      \centerline{Biomedical Imaging Group, École polytechnique fédérale de Lausanne}
		    \centerline{1015 Lausanne, Switzerland}
		} % Do not forget to end the {\footnotesize by the sign }

		\medskip

		\centerline{\scshape Michael Unser}
		\medskip
		{\footnotesize
            \centerline{michael.unser@epfl.ch}
		      \centerline{Biomedical Imaging Group, École polytechnique fédérale de Lausanne}
		    \centerline{1015 Lausanne, Switzerland}
		}

		\medskip

		\centerline{\scshape Jonathan Dong$^*$}
		\medskip
		{\footnotesize
            \centerline{jonathan.dong@epfl.ch}
		      \centerline{Biomedical Imaging Group, École polytechnique fédérale de Lausanne}
		    \centerline{1015 Lausanne, Switzerland}
		}

		\bigskip

        \vspace{1cm}
        
		\footnotetext[3]{The work is supported by European Union's Horizon Europe Research and Innovation Programme under Grant Agreement No. (101046667 (SWOPT)).}
        \footnotetext[4]{Corresponding author: 
        Jonathan Dong.
        }
    \end{center}

\begin{abstract} 
\vspace{0.5cm}

In this paper, we propose a global framework that includes a detailed model of the photo-switching and acoustic processes for photo-switching optoacoustic mesoscopy, based on the underlying physics.
We efficiently implement two forward models as matrix-free linear operators and join them as one forward operator.
Then, we reconstruct the concentration maps directly from the temporal series of acoustic signals through the resolution of one combined inverse problem.
For robustness against noise and clean unmixing results, we adopt a hybrid regularization technique composed of the $l_1$ and total-variation regularizers applied to two different spaces.
We use a proximal-gradient algorithm to solve the minimization problem.
Our numerical results show that our regularized one-step approach is the most robust in terms of noise and experimental setup.
It consistently achieves higher-quality images, as compared to two-step or unregularized methods.

\end{abstract}
\end{titlepage}

\section{Introduction}
\subsection{Background}
Optoacoustic (OA), also referred to as photoacoustic, imaging is a noninvasive multi-scale and multi-contrast imaging technology \cite{wang_practical_2016}.
It is widely adopted in biomedical research to study the anatomical, functional, molecular and metabolic aspects of living biological structures \cite{vu_listening_2019, yao_multiscale_2016}.
By capitalizing on the photoacoustic effect, OA imaging overcomes the strong scattering of photons in biological tissues and allows for deeper penetration and better resolution than traditional optical imaging \cite{ntziachristos_going_2010}.

OA imaging can be implemented in a variety of setups to accommodate for different imaging tasks \cite{wang_multiscale_2009}.
% One setup is tomography (OAT), in which the sample is fully or partially surrounded by the illumination and detection geometry.
% Widefield diffuse light and a detector array are used to achieve imaging with centimeter depth at a resolution of hundreds of microns \cite{wang_practical_2016}.
% Another setup is microscopy, in which a tightly focused illumination and a single-element transducer are coupled and  scan over the surface of the sample to directly form an image.
% In this case, the spatial resolution and penetration depth are comparable to that of an optical microscopy \cite{wang_practical_2016}.
In particular, OA mesoscopy (OAM) strikes a balance between spatial resolution and penetration depth, thus bridging the imaging gap between OA tomography and OA microscopy.
It uses loosely focused (or unfocused) diffuse illumination and a broadband detector to reach millimeter imaging depths at a resolution of tens of microns \cite{omar_optoacoustic_2019}.
The most common implementation of OAM is a raster-scanning system, in which a spherically focused single-element transducer with a large numerical aperture scans over an area on the surface of the sample \cite{he_fast_2022}.
The illumination is either coupled with the transducer to move across the surface of the sample \cite{omar_optoacoustic_2019, he_fast_2022} or from a fixed laser to avoid artifacts that come from the variations of the spatially dependent fluence \cite{aguirre_precision_2017}.
Although multi-element transducer arrays have been proposed to avoid the time-consuming scanning procedure \cite{omar_optoacoustic_2019}, they are difficult to manufacture without some sacrifice in sensitivity, center frequency, or bandwidth.

The contrast of OA imaging arises from the optical absorption of a wide variety of endogenous and exogenous molecules \cite{wang_photoacoustic_2012}.
% Endogenous ones such as hemoglobin, lipids and melanin enable label-free imaging \cite{wang_photoacoustic_2012}.
% Exogenous ones such as nanoparticles, organic dyes and reporter genes make OA suitable for targeted imaging as well \cite{wang_photoacoustic_2012, brunker_photoacoustic_2017, li_photoacoustic_2018}.
Among the exogenous contrast agents, a group of reversibly switchable protein reporters
% derived from a bacterial phytochrome
is an emerging choice to enhance the imaging sensitivity and specificity \cite{mishra_photocontrollable_2019}.
%\cite{brakemann_reversibly_2011, mark_dual-wavelength_2018, mishra_photocontrollable_2019, chee_vivo_2018, yao_multiscale_2016, li_small_2018}.
One drawback of the exogenous agents is low sensitivity in vivo because their signal is over-shadowed by the strongly absorbing endogenous chromophores such as hemoglobin \cite{wang_practical_2016, yao_multiscale_2016}.
Fortunately, photo-switching reporters offer a solution due to their special photo-physical property: their extinction profiles vary as they are illuminated by two different wavelengths (referred to as ON and OFF wavelengths) \cite{mishra_multiplexed_2020}. 
During photo-switching, the detected ultrasound signals are thus a temporal multiplex of photo-switching reporters and the unmodulated endogenous chromophores in the tissue.
This temporal multiplexing technique allows one to extract the signals of the introduced labels from the tissue background \cite{ mishra_multiplexed_2020}.

Photo-switching protein reporters combined with OA have demonstrated great potential in high-specificity multi-label imaging in the tomographic setup \cite{mishra_multiplexed_2020, li_small_2018, liu_model-based_2025}.
Yet, the integration of photo-switching with an OA mesoscopy setup is relatively new.
% Similar to photo-switching OAT, this technique promises further enhancement of the imaging specificity and sensitivity.
For it to be meaningful, it is vital to model the physical principles of photo-switching in the OAM setup and to develop a dedicated reconstruction and unmixing technique.
Then only will one be able to extract quantitative information of the reporters from the temporal series of OA signals.

\subsection{State of the Art}
Existing temporal unmixing methods in OAT proceed in two steps \cite{yao_multiscale_2016, stiel_high-contrast_2015, chee_vivo_2018, mishra_multiplexed_2020, li_small_2018, liu_model-based_2025}.
Based on existing algorithms for classic (non-photo-switching) OAT, 
they first solve independent acoustic inversion problems to obtain OA images for each switching pulse.
Then, they apply an unmixing algorithm on the stack of reconstructed OA images to recover spatial maps of protein species.

For the unmixing step, one usually focuses on the analysis of the OFF-switching series.
During an OFF-switching cycle, the OA signal is progressively decreased by a sequence of laser pulses at the OFF-wavelength.
The temporal evolution of the intensity of the OA signals at each spatial location approximately follows a decaying exponential model \cite{STANKEVYCH2021365, liu_model-based_2025}.
The speed of OFF-switching, characterized by the exponent parameter in the decay model, is the key to distinguishing different species and the background.

Differential imaging subtracts the last frame from the first of OA images of a cycle and works when there is only one species \cite{yao_multiscale_2016}.
Chee et al. extended it to the imaging of two species whose absorption spectra do not overlap \cite{chee_vivo_2018}.
Subsequent methods based on the fitting of an exponential model, followed by pixelwise classification, make better usage of the complete information of a cycle \cite{mishra_genetically_2021, STANKEVYCH2021365}.
Li et al. extended the decay model to include the local light-fluence intensity, an important factor that drives the switching speed \cite{li_small_2018}.
Such models have been refined by the inclusion of physical factors that play a role in the evolution of the OA signal,
which gives access to quantitative unmixing of multiple spatially overlapping species \cite{liu_model-based_2025}.

Regarding the acoustic inversion step, there have been extensive works on the model of the propagation of photoacoustic waves and the characterization of the detector. 
%\cite{xu_universal_2005, dean-ben_accurate_2012, ammari_mathematical_2012, scherzer_handbook_2011, wang_biomedical_2007, wang_photoacoustic_2009, liao_optoacoustic_2004, bai_improvement_2021, turner_improved_2014, li_improved_2006, hofmann_enhancing_2022}.
The wave equation describes the propagation of the acoustic waves originated from the OA sources in the sample  \cite{ammari_mathematical_2012, wang_biomedical_2007, scherzer_handbook_2011} under specific acoustic properties (for instance, acoustic attenuation and variable speed of sound) of the sample \cite{hristova_reconstruction_2008, ammari_mathematical_2012, scherzer_handbook_2011}.
In a few idealized scenarios, there exists an explicit back-projection-type inversion formula \cite{xu_universal_2005}.
In practice, one often makes reasonable assumptions on the acoustic properties of the medium such that the solution to the wave equation has an explicit expression, for instance, in the form of an integral over a sphere in 3D (or an arc in 2D) \cite{dean-ben_accurate_2012}.
% For instance, if one assumes the sample is acoustically homogeneous without attenuation, and the speed of sound is constant, the solution to the wave equation is a Poisson type integral that involves integration of the unknown optical energy function on a sphere in 3D (or an arc in 2D) and a first-order partial differentiation in time \cite{dean-ben_accurate_2012}.
This integral is further discretized into a linear system with a model matrix.
By solving it, one reconstructs the unknown optical energy map.
This model-based approach
has been refined to include the properties of the ultrasound transducer, the detection geometry \cite{dean-ben_accurate_2012, rosenthal_fast_2010, wang_imaging_2011}, and the fluence variation \cite{aguirre_low_2013} to improve the quality of the reconstruction.

In the context of OA mesoscopy, a popular alternative approach is the delay-and-sum algorithm.
Its advantage is speed and memory, as compared to model-based methods \cite{aguirre_low_2013}.
% Instead of the unfocused detection in tomography, one uses a spherically-focused transducer with a large numerical aperture to scan the surface of the sample and acquires a set of 1D raw acoustic signals.
Due to the limited depth-of-focus of the transducer, the quality of the image deteriorates significantly in the out-of-focus region.
The synthetic-aperture focusing technique (SAFT), adapted from ultrasound imaging \cite{li_improved_2006}, is used
to solve this issue.
% improve the resolution and reduce noise in the image.
It applies appropriate delays (relative to the acoustic focus) to the neighboring scan lines within the sensitivity range of the detector, then sums up the delayed signal to get rid of out-of-focus blur \cite{liao_optoacoustic_2004}.
Many variations of SAFT have been proposed; for instance, some that add correction factors and include the transducer properties (e.g., the electrical and spatial impulse response).
They have been shown to further improve the quality and signal-to-noise ratio of the OA images \cite{li_improved_2006, turner_improved_2014, bai_improvement_2021}.

In quantitative OAT without photo-switching, several researchers have concerned themselves with mathematical modeling and numerical simulations in the tomographic setting \cite{ding_one-step_2015, javaherian_direct_2019, haltmeier_single-stage_2015}.
There, methods to recover the unknown optical absorption maps from the acoustic measurements can be classified into two categories \cite{cox_quantitative_2012}.
The first one models the optical and acoustic processes individually, and then, solves two inverse problems (referred to as the two-step, or two-stage approach);
the methods of the second category join the forward operators of the two processes together as one composite operator and solve only one grand inverse problem (referred to as the one-step, or global, or single-stage approach) \cite{ding_one-step_2015, javaherian_direct_2019, haltmeier_single-stage_2015}.
Haltmeier et al. \cite{haltmeier_single-stage_2015} showed that the global approach improves the reconstruction quality, as compared to the two-step approach.

To the best of our knowledge, there has not been any work on the modeling and quantitative temporal unmixing algorithms of OA mesoscopy combined with photo-switching.

\subsection{Contribution}
In this paper, we present a mathematical framework that encapsulates a complete forward imaging model and a dedicated quantitative unmixing and global-reconstruction algorithm.
It focuses on a novel OAM setup that consists of widefield illumination from a fixed laser and an array of ultrasound transducers, combined with photo-switching protein reporters.

The full pipeline, from optical illumination to acoustic detection during an OFF-switching cycle, is the global forward operator.
It is itself composed of two operators: the optical model that includes the photo-switching responses; and the acoustic model.
The optical model, based on our previous work \cite{liu_model-based_2025}, offers a detailed description of the temporal evolution of the signal during photo-switching and includes the impact of local light fluence and of the intrinsic kinetics of the reporters.
The acoustic model follows the the principles of SAFT and includes the properties of the transducer through spatial integration with the spatial response of the transducer on the wavefront and temporal correlation with the electrical response of the transducer.
On the computational aspect of our approach, the acoustic forward model is constructed as a matrix-free linear operator and implemented efficiently, in a way that avoids the computational bottlenecks that existing model-based approaches do face.

Then, we follow the model-based approach and formulate the inverse problem of the recovery of the spatial concentration maps from the acoustic measurements as a minimization problem in which we incorporate prior information in the form of sparsity-promoting regularization.
We solve the global inverse problem using a proximal-gradient-based iterative algorithm.

We validate our framework on numerical simulations and show the performance of our proposed regularized global unmixing method.
Finally, we explain the implementation of our models and carry out a computational analysis and speed benchmark.

\section{Methods}
\subsection{Forward Pipeline}
\begin{figure}[tb]
    \centering
    \includegraphics[width=\linewidth]{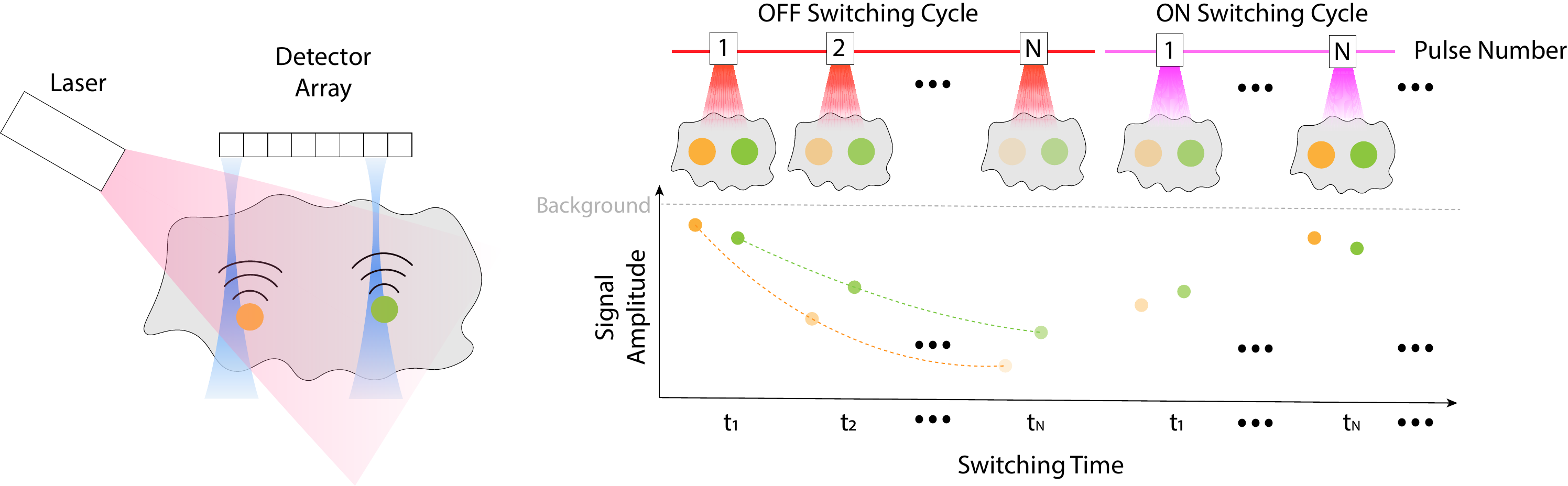}
    \caption{(Left) Experimental setup.
    The sample being imaged is represented by the gray object.
    The orange and green discs represent two species of photo-switching reporters.
    The pink area represents the diffuse illumination from the laser.
    The sensitivity field of two arbitrary transducers in the array of detectors is depicted by the blue areas.
    (Right) Principle of photo-switching.
    The OFF and ON switching cycles (also wavelengths used in the cycle) are indicated by color red and magenta, respectively.
    On top, the numbers on the lasers represent the pulse number within a cycle.
    On the bottom, $t_1, t_2, \ldots, t_N$ represent the discrete time points during a switching cycle.
    Dashed curves with the same color-code as the reporters during the OFF cycle illustrate the evolution of the amplitude of the OA signal.
    The dashed gray horizontal line indicates the evolution of a point in the background.
    }
    \label{fig:principle}
\end{figure}

\subsubsection{Imaging Principle}
Photo-switching OAM relies on a scheduled illumination of ON and OFF switching pulses.
% , after each pulse, a complete OAM process from illumination to detection occurs to form one measurement image.
Each laser pulse gives rise to a complete OA process. 
After the surface of the tissue is illuminated, photons are absorbed and scattered by the tissue, which creates a fluence field.
As photons propagate through the tissue, chromophores absorb some of the optical energy that is converted to heat, leading to a thermal expansion and local rise in pressure.
The change of pressure propagates as ultrasonic waves that are detected by a linear transducer array at the surface of the tissue.
The measured acoustic signals are used to reconstruct the original deposition of optical energy and other optical properties of interest, for instance, the absorption coefficient.

The illumination schedule consists of ON and OFF cycles, each containing a sequence of laser pulses.
Within a cycle, the pulses lead the protein molecules of all the species to transit stochastically from one state into the other.
As a result, the extinction parameters of these species progressively shift from one state to the other.
The generated OA signal, which is a sum of the contribution from all the species and the unmodulated background thus exhibits an evolution over the switching time.
The conversion between these two molecular states of the protein is reversible and impervious to photo-fatigue, which allows one to assume that the concentration of each species is constant over time.
Typically, the signal evolution during the OFF cycles is preferred for analysis as the quality of the signal is better than the ON cycles.

\begin{figure}[t!]
    \centering
    \includegraphics[width=0.9\linewidth]{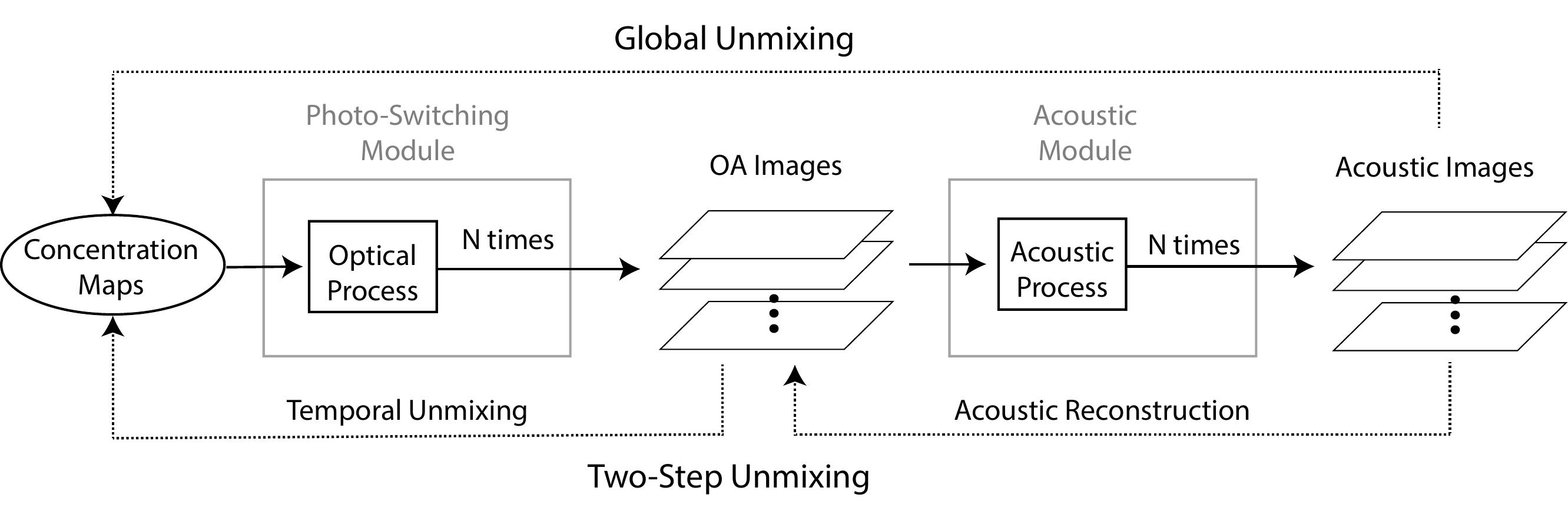}
    \caption{Forward pipeline and inversion approaches.}
    \label{fig:full-pipeline}
\end{figure}

The measurements hence consist of the collection of the detected acoustic signals for each pulse, from which one can  recover the maps of the spatial distribution of each species.
The setup and principle of photo-switching OA mesoscopy are shown in Figure \ref{fig:principle}, while Figure \ref{fig:full-pipeline} contains the complete forward pipeline and the two approaches of temporal unmixing.

\subsubsection{Photo-Switching Model}
We carry out our modeling on a 2D sample during one OFF-switching cycle.
We define the sample to be imaged as a function of spatial locations $\mathbf{r}=(x, z)\in\mathbb{R}^2$ with a compact support $\Omega\subset\mathbb{R}^2$.

The forward model of photo-switching has been derived in our previous work \cite{liu_model-based_2025}.
Here, we briefly recall its key ingredients.
During an OFF-switching cycle, we assume that the pulses exhibit no mutual dependence and that there is no temporal overlap between neighboring switching events.
We start by modeling the evolution of the extinction coefficient $\varepsilon(\mathbf{r}, t)$ of a reporter at switching time $t$ with the exponential law
\begin{equation}
    \label{eq:epsilon}
    \varepsilon(\mathbf{r}, t) =
    \overline{\varepsilon}
    \text{e}^{-k\Phi(\mathbf{r})t} + \varepsilon^{\text{OFF}}
\end{equation}
where $\overline{\varepsilon} = (\varepsilon^{\text{ON}} - \varepsilon^{\text{OFF}})$ is the difference between the extinction coefficients at the OFF-switching wavelength of the ON and OFF states, $k>0$ is the kinetic constant,
and $\Phi(\mathbf{r})$ is the distribution of light fluence. 
The parameters $\varepsilon^{\text{ON}}$, $ \varepsilon^{\text{OFF}}$, and $k$ can be determined experimentally.

The speed of switching, indicated by the exponent parameter, is influenced by both the intrinsic kinetics of the protein and the local fluence intensity.
We make the assumption that the spatially varying light fluence $\Phi(\mathbf{r})$ does not depend on the switching time as we assume that the contribution of the background (e.g., blood vessels) is much stronger than that of the protein reporters.

The sum of the extinction of all the species weighted by their respective concentration $c(\mathbf{r})$, in addition to the background, yields the total absorption map
\begin{equation}
\label{eq:mu_a}
    \mu_\text{a}(\mathbf{r}, t) = \sum_{p=1}^{P}
    \varepsilon_p(\mathbf{r}, t)
    c_p(\mathbf{r}).
\end{equation}
Here, we consider the general case of $(P-1)$ species and model the contribution of the background as the last ($P$th) reporter such that $\varepsilon_P(\mathbf{r}) c_P(\mathbf{r})=\mu_{\text{a}}^{\text{bg}}(\mathbf{r})$, where $\mu_{\text{a}}^{\text{bg}}(\mathbf{r})$ is the absorption map of the tissue background.
The fluence field and absorption map jointly give rise to the deposited optical energy $H(\mathbf{r}, t)$ at switching time $t$, as
\begin{equation}\label{eq:final-H}
    H(\mathbf{r}, t) = 
    \Phi(\mathbf{r})\sum_{p=1}^{P}
    \left(\overline{\varepsilon}_p
    \e^{-k_p\Phi(\mathbf{r})t} +
    \varepsilon_p^{\text{OFF}}\right)
    c_p(\mathbf{r}),\quad n=0, \ldots, N-1,
\end{equation}
where the quantity of interest $c_p(\mathbf{r})$ is the spatial distribution map of the $p$th species.
The non-switching background is included in \eqref{eq:final-H} as the $P$th reporter, with $k_P=0$ and $\overline{\varepsilon}_P=0$. 

\subsubsection{Optical Model}
The spatial distribution of the light fluence is governed by the absorption and scattering of the photons inside the tissue.
We precompute the fluence map $\Phi^0(\mathbf{r})$ without the contribution of the reporters and assume that $\Phi(\mathbf{r}) \approx \Phi^0(\mathbf{r})$.
This assumption is reasonable, owing to the small contribution of the photo-switching reporters to the absorption map.
Due to the diffuse nature of the illumination and the millimeter penetration depth, the photon propagation in our setup satisfies the assumption that the scattering of photons is much stronger than the absorption.
Hence, we use the diffusion equation \cite{lorenzo_principles_2012}, a first-order approximation to the radiative transfer equation (RTE), to compute the fluence map.
While RTE is accurate, it is difficult to solve \cite{ammari_multi-wave_2017} and its numerical equivalence, the Monte Carlo method, is computationally expensive, too \cite{wang_biomedical_2007}.
We adopt instead the diffusion equation \eqref{eq:de}, complemented by the Robin-type boundary condition \eqref{eq:bc} (for the case of tissue-water interface), to predict the fluence $\Phi(\mathbf{r})$ as
\begin{eqnarray}
    \label{eq:de}
    \mu_{\text{a}}(\mathbf{r})\Phi(\mathbf{r}) - \nabla\cdot (D(\mathbf{r})\nabla\Phi(\mathbf{r})) =& I(\mathbf{r}), \qquad & \mathbf{r} \in \Omega, \\
    \label{eq:bc}
    \Phi(\mathbf{r}) - 2D(\mathbf{r})\nabla \Phi(\mathbf{r})\cdot\mathbf{n} =& 0, \qquad & \mathbf{r} \in \partial \Omega
\end{eqnarray}
where $\partial\Omega$ represents the boundary of the sample and $\mathbf{n}$ is the outward normal vector on the boundary.
As the illumination is wide-field and can be assumed to be homogeneous when it reaches the surface of the sample, we model it with a function $I(\mathbf{r})$.
It describes a line segment with center $\mathbf{r}_\text{c} = (x_\text{c}, 0)$, of length $W>0$ and uniform intensity $I_0>0$ 
\begin{equation}
    \label{eq:light-source}
    I(\mathbf{r}) = \left\{
        \begin{array}{ll}
        I_0, & \text{if } |x-x_\text{c}|<\frac{W}{2}\\
        0, & \text{else}.
        \end{array}
    \right.
\end{equation}
The spatially varying diffusion coefficient map $D(\mathbf{r})$ depends on the absorption coefficient map $\mu_{\text{a}}(\mathbf{r})$, the anisotropy factor $g\in(0, 1)$ (which we set to a typical value of 0.9), and the scattering coefficient map $\mu_{\text{s}}(\mathbf{r})$ according to \cite{ammari_multi-wave_2017}
\begin{equation}
    \label{eq:diffusion-coefficient}
    D(\mathbf{r})=\frac{1}{3(\mu_{\text{a}}(\mathbf{r}) + (1-g)\mu_{\text{s}}(\mathbf{r}))}.
\end{equation}
We set $\mu_{\text{a}}(\mathbf{r})$ and $\mu_{\text{s}}(\mathbf{r})$ the same as the counterparts of the non-switching background because we assume the contribution from the reporters to be negligible.
By solving \eqref{eq:de} and \eqref{eq:bc}, we obtain a map $\Phi^0(\mathbf{r})$ of the fluence distribution within the sample and use it to construct the forward model \eqref{eq:final-H} for photo-switching.

\subsubsection{Acoustic Model}
The deposited optical energy $H$ absorbed by the tissue leads to a local rise in temperature.
It causes a thermo-elastic expansion of the tissue and produces acoustic waves that propagate through the sample.
The detector on the sample surface records the photoacoustic waves as measurements.

Because the acoustic propagation and detection is independent for each pulse and its mathematical model takes the same form, we therefrom omit the notation $t$ in the modeling of a single acoustic process, for the sake of simplicity.
The generated initial pressure (acoustic) field $b(\mathbf{r})$ is proportional to the optical energy such that
\begin{equation}
\label{eq:pa-effect}
    b(\mathbf{r}) = \iota(\mathbf{r}) H(\mathbf{r}),
\end{equation}
where $\iota(\mathbf{r})$ is the unitless Grueneisen coefficient.
It indicates the efficiency of conversion between heat and pressure, and we assume it to be constant and set its value to 1 for simplicity \cite{haltmeier_single-stage_2015, javaherian_direct_2019}.

\begin{figure}[t]
    \centering
    \includegraphics[width=0.8\linewidth]{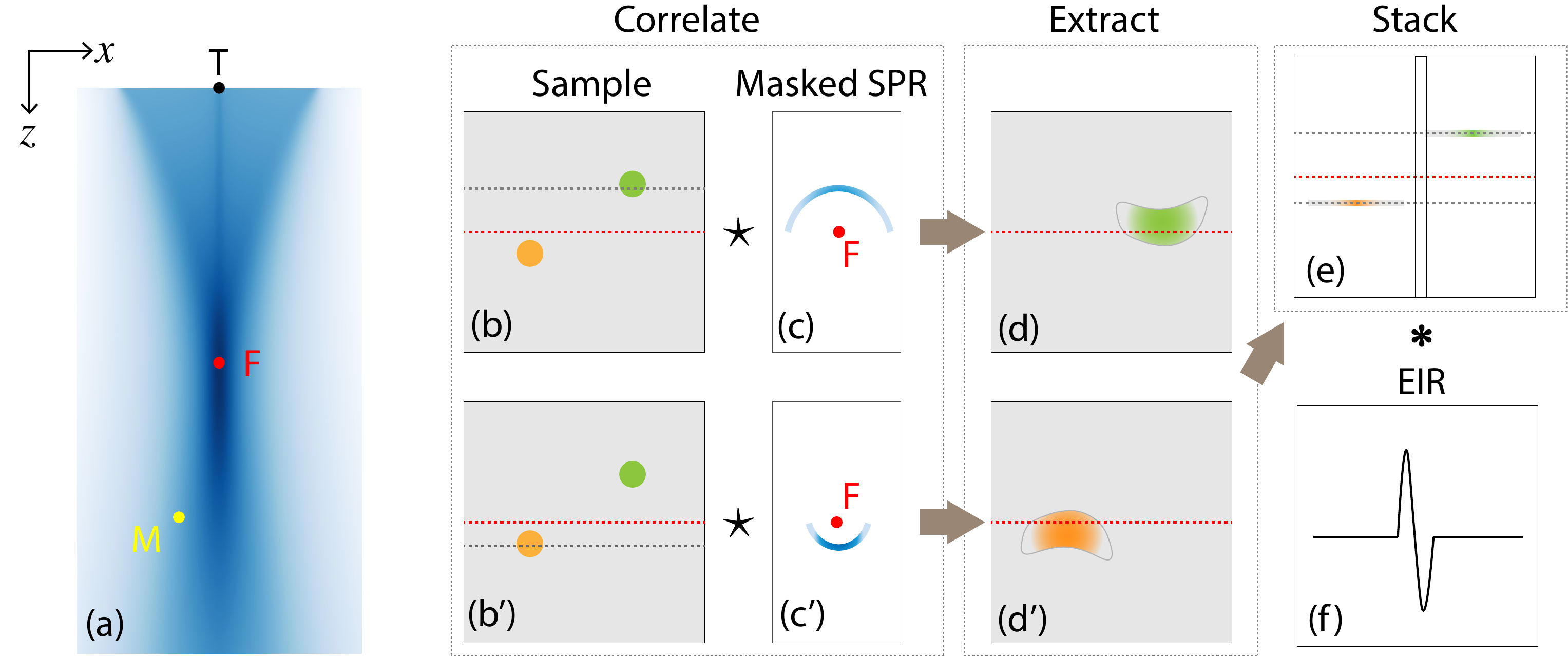}
    \caption{
    (a) Spatial response of the transducer.
    The blue area represents the sensitivity field of the transducer.
    The dots represent the center of the transducer (T), the focal spot (F) and a point OA source (M).
    (b)-(f) Generation of the acoustic signals.
    SPR: spatial response of the transducer.
    EIR: electrical impulse response of the transducer.
    (b) and (b'): Sample.
    The background is in gray, the green and orange discs represent two photo-switching reporters.
    The dashed horizontal red line represents the location of the focal plane.
    The dashed gray lines in (b) and (b') indicate the horizontal line of pixels of interest, one above (b) and one below (b') the focal plane.
    (c) and (c'): Masked SPR (map of the weighted curves that corresponds to the depth of interest in (b) and (b'), respectively).
    (d) and (d'): Correlation between the sample and the masked SPR (c) and (c').
    The focal plane (red dashed line) indicates where to extract the line of pixels.
    (e) Spatial integration step by stacking the extracted lines at the corresponding locations indicated in (b) and (b').
    The narrow vertical box indicates a line of pixels on which we convolve with the EIR of the transducer (f).
    }
    \label{fig:acoustic-signal-generation}
\end{figure}

During one acoustic process, the ultrasonic waves originating from sources within the sample propagate through the sample and are detected by the transducer array on the surface of the sample.
We assume a constant speed of sound $c_0$ and an acoustically homogeneous and non-attenuating medium.
We denote by $\tau$ the acoustic propagation time (microsecond scale, propagation of sound over a few millimeters) to distinguish it from the switching time $t$ (millisecond scale, repetition rate of the laser).

Following the principles of SAFT, we assume
that the signal from the focal point arrives at the same time on all the area of the transducer element, owing to its parabolic geometry, 
Therefore, we use the focal spot as a reference and propagate back to it.
We denote the transducer position by $\mathbf{r}_{\text{T}}=(x_{\text{T}}, 0)$, the focal length by a constant $f>0$, and the location of the focal spot as $\mathbf{r}_{\text{F}}=(x_{\text{T}}, z_{\text{F}})$.
The detected signal on the transducer at time $\tau$, comes from the contributions of all the point sources lying on the (upper) semicircle $C_1^{(x_{\text{T}}, \tau)} = \left\{ \mathbf{r} : |\mathbf{r} - \mathbf{r}_{\text{F}}| = f - c_0\tau, z<z_{\text{F}} \right\}$ centered at $\mathbf{r}_{\text{F}}$ with radius $(f-c_0\tau)$ when $c_0\tau < f$ , and on another (lower) semicircle $C_2^{(x_{\text{T}}, \tau)} = \left\{ \mathbf{r} : |\mathbf{r} - \mathbf{r}_{\text{F}}| = c_0\tau - f, z>z_{\text{F}} \right\}$ when $c_0\tau > f$.
At the focal spot $\mathbf{r}_\text{F}$, i.e., when $c_0\tau=f$, $q(\mathbf{r}_\text{F})$ is defined directly as $b(\mathbf{r}_\text{F})$.
Further, we take into account the influence of the sensitivity of the transducer, described by its (shifted) spatial response function $h(\mathbf{r} - \mathbf{r}_{\text{T}})$ at the location $\mathbf{r}$ of the point source (see Figure \ref{fig:acoustic-signal-generation} (a)).
Hence, we integrate the weighted amplitude $b(\mathbf{r})h(\mathbf{r} - \mathbf{r}_{\text{T}})$ on $C_1^{(x_{\text{T}}, \tau)}$ when $c_0\tau < f$
\begin{equation}
\label{eq:q-above-1d}
    q(x_{\text{T}}, \tau) = \int_{C_1^{(x_{\text{T}}, \tau)}} b(\mathbf{r})h(\mathbf{r} - \mathbf{r}_{\text{T}}) \text{d}s,
\end{equation}
and on $C_2^{(x_{\text{T}}, \tau)}$ when $c_0\tau > f$
\begin{equation}
\label{eq:q-below-1d}
    q(x_{\text{T}}, \tau) = \int_{C_2^{(x_{\text{T}}, \tau)}} b(\mathbf{r})h(\mathbf{r} - \mathbf{r}_{\text{T}}) \text{d}s,
\end{equation}
where $\text{d}s$ is the 1D arc-length element.
We denote the radius of the semicircle as $r_\tau=|f-c_0\tau|$ for the sake of simplicity. 
By representing the 1D line integral as a constrained 2D integral via a Dirac delta function, we formulate equation \eqref{eq:q-above-1d} equivalently as
\begin{equation}
\label{eq:q-above-2d}
    q(x_{\text{T}}, \tau) = 
    \int_{\mathbb{R}}\text{d}x \int_{-\infty}^{z_{\text{F}}}\text{d}z \; b(x, z)  h(x - x_{\text{T}}, z - z_{\text{F}})  \delta\left((x - x_{\text{T}})^2 + (z - z_{\text{F}})^2 - r_\tau^2 \right) (2r_\tau),
\end{equation}
where $2r_\tau$ represents the Jacobian correction to account for the change of variables in the Dirac delta function.
Similarly for equation \eqref{eq:q-below-1d}, we obtain
\begin{equation}
    \label{eq:q-below-2d}
    q(x_{\text{T}}, \tau) = 
    \int_{\mathbb{R}}\text{d}x \int_{z_{\text{F}}}^{\infty}\text{d}z \; b(x, z) b(x, z)  h(x - x_{\text{T}}, z - z_{\text{F}}) \delta\left((x - x_{\text{T}})^2 + (z - z_{\text{F}})^2 - r_\tau^2 \right) (2r_\tau).
\end{equation}
Next, we take into account the electrical impulse response $v(\tau)$ of the transducer and model its effect as the temporal convolution of $q(x_{\text{T}}, \tau)$ with $v(\tau)$
\begin{equation}
    \label{eq:acoustic-temporal-conv}
    p(x_{\text{T}}, \tau) = \int_{\mathbb{R}} q(x_{\text{T}}, \tau')  v(\tau - \tau') \text{d}\tau',
\end{equation}
where $p(x_{\text{T}}, \tau)$ is the signal detected on the transducer.

\subsection{Numerical Implementation}
We discretize the 2D object domain $\Omega$ into a collection of points $\Delta\boldsymbol{i} = (\Delta_1 i_1, \Delta_2 i_2)$, where $\boldsymbol{i} = (i_1, i_2) \in \Omega_{\text{2D}}\subset \mathbb{Z}^2$ is the index of the sampled points,
$i_1 = 0, \ldots, L_\text{x} - 1$, and $i_2=0, \ldots, L_\text{z} - 1$.
The diagonal matrix $\Delta = \text{diag}\left\{\Delta_1, \Delta_2\right\}$ defines the stepsize $\Delta_1$ and $\Delta_2$ for the lateral and axial direction, respectively.
The switching time $t$ are sampled at $N$ pulses such that $t_n = \Delta_{\text{t}} n$, $n=0, \ldots, N-1$, with $\Delta_{\text{t}}$ the temporal stepsize which is determined by the laser repetition rate during switching.

\subsubsection{Photo-Switching Model}
We define the discrete energy deposition $H_{\boldsymbol{i}}^{n}= H(\mathbf{r}_{i_1, i_2},t_{n} )$, the reporter concentrations $c^p_{\boldsymbol{i}} = c^{p} (\mathbf{r}_{i_1,i_2} )$, $p=1, \ldots, P$ with $P$ the total number of species plus one (the background), and fluence distribution $\Phi_{\boldsymbol{i}} = \Phi (\mathbf{r}_{i_1,i_2} )$. 
The photo-switching forward model \eqref{eq:final-H} is first specified at a pixel index $\boldsymbol{i}$ for all $N$ time points as
\begin{equation}
    \label{eq:Sij-detailed}
    \underbrace{
    \begin{bmatrix}
    H^{0}_{\boldsymbol{i}} \\ H^{2}_{\boldsymbol{i}} \\ \vdots \\ H^{N-1}_{\boldsymbol{i}} 
    \end{bmatrix}}
    _{\mathbf{H}_{\boldsymbol{i}}}
    = 
    \underbrace{
    \hat\Phi_{\boldsymbol{i}}
    \begin{bmatrix}
    \bar{\varepsilon}_{1} \e^{-k_{1} \Phi_{\boldsymbol{i}} t_{0}} + \varepsilon_{1}^{\text{OFF}}
    &  
    \cdots & 
    \bar{\varepsilon}_{P} \e^{-k_{P} \Phi_{\boldsymbol{i}} t_{0}} + \varepsilon_{P}^{\text{OFF}}\\
    \bar{\varepsilon}_{1} \e^{-k_{1} \Phi_{\boldsymbol{i}} t_{1}} + \varepsilon_{1}^{\text{OFF}}
    & 
    \cdots & 
    \bar{\varepsilon}_{P} \e^{-k_{P} \Phi_{\boldsymbol{i}} t_{1}} + \varepsilon_{P}^{\text{OFF}}\\
    \vdots &  
    \ddots & 
    \vdots \\
    \bar{\varepsilon}_{1} \e^{-k_{1} \Phi_{\boldsymbol{i}} t_{N-1}} + \varepsilon_{1}^{\text{OFF}}
    & 
    \cdots & 
    \bar{\varepsilon}_{P} \e^{-k_{P} \Phi_{\boldsymbol{i}} t_{N-1}} + \varepsilon_{P}^{\text{OFF}}
    \end{bmatrix}}
    _{\mathbf{S}_{\boldsymbol{i}}}
    \underbrace{
    \begin{bmatrix}
    c^1_{\boldsymbol{i}} \\ c^2_{\boldsymbol{i}} \\ \vdots \\ c^P_{\boldsymbol{i}} 
    \end{bmatrix}}
    _{\mathbf{c}_{\boldsymbol{i}}},
\end{equation}
where $\mathbf{H}_{\boldsymbol{i}} \in \mathbb{R}^{N}$, $\mathbf{c}_{\boldsymbol{i}} \in \mathbb{R}^{P}$, and $\mathbf{S}_{\boldsymbol{i}} \in \mathbb{R}^{N \times P}$. 
Then, we assemble the per-pixel system of equation \eqref{eq:Sij-detailed} into a block-diagonal system by sequentially combining $L=L_{\text{x}} L_{\text{z}}$ systems 
%-
\begin{equation}
    \label{eq:photoswitching-full-discrete-forward-detailed}
    \underbrace{
    \begin{bmatrix}
    \mathbf{H}_{0,0} \\ \mathbf{H}_{1,0} \\ \vdots \\
    \mathbf{H}_{L_{\text{x}}-1,0} \\ \mathbf{H}_{0,1} \\ \vdots \\
    \mathbf{H}_{L_{\text{x}}-1,L_{\text{z}}-1}
    \end{bmatrix}}
    _{\mathbf{H}}
    = 
    \underbrace{
    \begin{bmatrix}
    \mathbf{S}_{0,0} & 
    \mathbf{0} &     
    \cdots & 
    \cdots & 
    \cdots & 
    \cdots & 
    \mathbf{0} \\
    \mathbf{0} & 
    \mathbf{S}_{1,0} &
    \ddots &     
    \cdots & 
    \cdots & 
    \cdots & 
    \vdots \\
    \vdots & 
    \ddots &
    \ddots &     
    \mathbf{0} & 
    \cdots & 
    \cdots & 
    \vdots \\
    \vdots & 
    \cdots & 
    \mathbf{0} &     
    \mathbf{S}_{L_{\text{x}}-1,0} &
    \mathbf{0} & 
    \cdots & 
    \vdots \\
    \vdots & 
    \cdots & 
    \cdots & 
    \mathbf{0} &     
    \mathbf{S}_{0,1} &
    \ddots & 
    \vdots \\
    \vdots & 
    \cdots & 
    \cdots & 
    \cdots & 
    \ddots &     
    \ddots &
    \mathbf{0} \\
    \mathbf{0} & 
    \cdots & 
    \cdots & 
    \cdots & 
    \cdots & 
    \mathbf{0} &     
    \mathbf{S}_{L_{\text{x}}-1, L_{\text{z}}-1} \\
    \end{bmatrix}}
    _{\mathbf{S}}
    \underbrace{
    \begin{bmatrix}
    \mathbf{c}_{0,0} \\ \mathbf{c}_{1,0} \\ \vdots \\
    \mathbf{c}_{L_{\text{x}}-1,0} \\ \mathbf{c}_{0,1} \\ \vdots \\
    \mathbf{c}_{L_{\text{x}}-1,L_{\text{z}}-1}
    \end{bmatrix}}
    _{\mathbf{c}},
\end{equation}
with the deposited energy vector $\mathbf{H} \in \mathbb{R}^{NL}$, concentration maps $\mathbf{c} \in \mathbb{R}^{PL}$, and the pooled system matrix $\mathbf{S} \in \mathbb{R}^{NL \times PL}$.
For the sake of memory efficiency, the block matrix $\mathbf{S}_{\boldsymbol{i}}$ for all the locations is directly constructed using Einstein summation\footnote{https://numpy.org/doc/stable/reference/generated/numpy.einsum.html} and the final forward matrix $\mathbf{S}$ is implemented as a block-diagonal operator without storing the zeros in it.

\subsubsection{Fluence}
We compute the discrete fluence map used to constructed the forward matrix by applying the finite element method to \eqref{eq:de} and \eqref{eq:bc} \cite{liu_model-based_2025}.
We briefly summarize the key steps on the variational formulation of the diffusion equation.
First, we multiply \eqref{eq:de} with a test function $v(\mathbf{r})\in H^1(\Omega)$ and integrate over $\Omega$ to obtain
\begin{equation}
    \label{eq:de-multiplied-with-v}
    \int_{\Omega}\mu_{\text{a}}(\mathbf{r})\Phi(\mathbf{r}) v(\mathbf{r}) \d\r-\int_{\Omega} \nabla\cdot (D(\mathbf{r})\nabla\Phi(\mathbf{r}))v(\mathbf{r})\d\r =\int_{\Omega} I(\mathbf{r})v(\mathbf{r})\d \r,
\end{equation}
where $H^1(\Omega)$ is a Sobolev space that contains square-integrable functions with square-integrable weak derivatives on $\Omega$, and $\d\mathbf{r}$ is the differential element on $\Omega$.
Then, we integrate by parts to get
\begin{equation}
    \label{eq:integration-by-parts}
    \int_{\Omega}\mu_{\text{a}}(\mathbf{r})\Phi(\mathbf{r}) v(\mathbf{r}) \d\r+\int_{\Omega} D(\mathbf{r})\nabla\Phi(\mathbf{r})\cdot\nabla v(\mathbf{r})\d\r-\int_{\partial\Omega}(D(\mathbf{r})\underset{\frac{\Phi(\mathbf{r})}{2D(\mathbf{r})}}{\underbrace{\nabla\Phi(\mathbf{r})\cdot\mathbf{n}})}v(\mathbf{r})\d s =\int_{\Omega}I(\mathbf{r})v(\mathbf{r})\d \r,
\end{equation}
where $\d s$ denotes differential element on the boundary of the domain.
Finally, we reorganize the terms and conclude with
\begin{equation}
    \label{eq:final-pde}
    \int_{\Omega}(\left(\mu_{\text{a}}(\mathbf{r})\Phi(\mathbf{r}) - I(\mathbf{r})\right) v(\mathbf{r})+ D(\mathbf{r})\nabla\Phi(\mathbf{r})\cdot\nabla v(\mathbf{r}))\d\mathbf{r} = \int_{\partial\Omega}\frac{\Phi(\mathbf{r})v(\mathbf{r})}{2}\d s.
\end{equation}
Equation \eqref{eq:final-pde} is implemented and solved using Fenicsx, an open-source library for the numerical solution of partial differential equations \cite{LoggEtal_11_2012, AlnaesEtal2015}.
We provide more details in Appendix A.
The solution on the finite element mesh is projected to the Cartesian grid via linear interpolation.

\subsubsection{Acoustic Model}
The implementation of the acoustic model lies in the discretization of \eqref{eq:q-above-2d} and \eqref{eq:q-below-2d}, where  $q(x_{\text{T}}, \tau)$ depends on both space and time.
Here, we introduce a convenient variable $y = c_0\tau$ and express $q(x_{\text{T}}, \tau)$ equivalently in pure spatial coordinates as $q(x_{\text{T}}, y)$ via a change of variables.
We discretize the measurement domain  into a grid of pixels $\Gamma\boldsymbol{m}$, where $\boldsymbol{m}=(m_1, m_2) \in \Theta_{\text{2D}}\subset\mathbb{Z}_2$ is the pixel index,
$m_1=0, \ldots, M_1 - 1$, and $m_2=0, \ldots, M_2 - 1$.
The set of indices $\Theta_{\text{2D}}$ has $M = M_1M_2$ elements.
The diagonal matrix $\Gamma = \text{diag}(\gamma_1, \gamma_2)$ contains the sampling stepsize $\gamma_1$ and $\gamma_2$ for the lateral and axial direction, respectively.

We define the forward model described in equation \eqref{eq:q-above-2d} as a linear operator $\mathcal{L}:L_2(\mathbb{R}^2)\to L_2(\mathbb{R}^2)$ and write \eqref{eq:q-above-2d} equivalently as
\begin{equation}
\label{eq:q-above-compact}
    q(x_{\text{T}}, y) = \mathcal{L}\{b\}(x_{\text{T}}, y).
\end{equation}
Similarly, the discretized measurement function $q:\Theta_{\text{2D}}\to\mathbb{R}$ can be written as
\begin{equation}
\label{eq:q-above-compact-discrete}
    q[\boldsymbol{m}] = \mathcal{L}\{b\}
    (\Gamma\boldsymbol{m}).
\end{equation}
We represent the compactly supported function $b(\mathbf{r})$ (equivalently $H(\mathbf{r})$, c.f \eqref{eq:pa-effect})
via a series of shifted basis functions $\varphi$
\begin{equation}
\label{eq:b-on-basis}
    b(\mathbf{r})
    = \sum_{\boldsymbol{k}\in \Omega_{\text{2D}}} \beta[\boldsymbol{k}] \varphi(\mathbf{r} / \Delta - \boldsymbol{k}), \quad \forall \mathbf{r}\in \Omega.
\end{equation}
Here, we choose the basis function $\varphi(\mathbf{r})$ to be the 2D rectangular function $\text{rect}(\mathbf{r})$ to represent the regular pixel grid and
\begin{equation}
    \text{rect}(\mathbf{r}) = \left\{
    \begin{aligned}
        1, & \quad -\frac{1}{2}\leq x \leq \frac{1}{2} \text{ and } -\frac{1}{2}\leq z \leq \frac{1}{2}, \\
    0, & \quad \text{ else}.
    \end{aligned}\right.
\end{equation}
Thanks to the linearity of $\mathcal{L}$, we obtain
\begin{equation}
\label{eq:q-above-on-basis}
        q[\boldsymbol{m}] = \sum_{\boldsymbol{k}\in \Omega_{\text{2D}}}\beta[\boldsymbol{k}] \mathcal{L}\{\varphi(\cdot / \Delta -\boldsymbol{k})\}(\Gamma\boldsymbol{m}),
\end{equation}
where the quantity $\mathcal{L}\{\varphi(\cdot / \Delta -\boldsymbol{k})\}(\Gamma\boldsymbol{m})$ has the expression
\begin{equation}
\label{eq:L-above}
     \mathcal{L}\{\varphi(\cdot / \Delta -\boldsymbol{k})\}(\Gamma\boldsymbol{m}) =  \int_{I_{\text{x}}}\int_{I_{\text{z}}} h(\mathbf{r} - \mathbf{r}_T) \delta\left((x - x_{\text{T}})^2 + (z - z_{\text{F}})^2 -r_\tau^2 \right) (2r_\tau)\text{d}z\text{d}x,
\end{equation}
with $I_{\text{x}} = [(k_1 - 1/2)\Delta_1, (k_1 + 1/2)\Delta_1]$, and $I_{\text{z}} = [(k_2 - 1/2)\Delta_2, (k_2 + 1/2)\Delta_2]$.

To compute a discrete version of this integral, we express the function $h(\mathbf{r})$ on the same rectangular basis with the same grid and support as $b(\mathbf{r})$:
\begin{equation}
\label{eq:h_T-on-basis}
    h(\mathbf{r}) = \sum_{\boldsymbol{k}\in \Omega_{\text{2D}}} \eta[\boldsymbol{k}] \text{rect}(\mathbf{r} / \Delta -\boldsymbol{k}), \quad \forall \mathbf{r}\in \Omega.
\end{equation}
We also assume that the discretization grids $\Omega_{\text{2D}}$ and $\Theta_{\text{2D}}$ match and that $\boldsymbol{r}_{\text{T}} = (x_\text{T},  z_\text{F}) = \Delta \boldsymbol{k}_{\text{T}}$ with $\boldsymbol{k}_{\text{T}} \in \mathbb{Z}^2$ the indices of the transducer focal position. 
Then,
\begin{equation}
\label{eq:h_T-on-basis}
    h(\mathbf{r} - \boldsymbol{r}_{\text{T}}) = \sum_{\boldsymbol{k}\in \Omega_{\text{2D}}} \eta[\boldsymbol{k} - \boldsymbol{k}_{\text{T}}] \text{rect}(\mathbf{r} / \Delta -\boldsymbol{k}), \quad \forall \mathbf{r}\in \Omega.
\end{equation}

In Eq.~\ref{eq:L-above}, $h$ is constant in the support $I_\text{x} \times I_\text{z}$, such that:
\begin{equation}
\label{eq:L-above2}
     \mathcal{L}\{\varphi(\cdot-\boldsymbol{k})\}(\Gamma\boldsymbol{m}) =  \eta[\boldsymbol{k}-\boldsymbol{k}_\text{T}] \int_{I_{\text{x}}}\int_{I_{\text{z}}} \delta\left((x - x_{\text{T}})^2 + (z - z_{\text{F}})^2 - r_\tau^2 \right) (2r_\tau)\text{d}z\text{d}x,
\end{equation}
The integral in equation \eqref{eq:L-above2}
reduces to computing the arc length of the semicircle intersecting with the 2D box defined by $(I_{\text{x}}\times I_{\text{z}})$, which we denote by $ s_{\boldsymbol{m}}[\boldsymbol{k}]$.
Therefore, \eqref{eq:q-above-on-basis} becomes
\begin{equation}
\label{eq:q-above-simplified}
    q[\boldsymbol{m}] = \sum_{\boldsymbol{k}\in \Omega_{\text{2D}}} \beta[\boldsymbol{k}] \eta[\boldsymbol{k} - \boldsymbol{k}_{\text{T}}]  s_{\boldsymbol{m}}[\boldsymbol{k}].
\end{equation}

The precise computation of $ s_{\boldsymbol{m}}[\boldsymbol{k}]$ for all spatial locations is  expensive, especially for large-scale reconstructions. 
For computational speed, we use a unit length of 1 for all pixels that intersect with the semicircle. 
Equation \eqref{eq:q-above-simplified} simplifies to 
\begin{equation}
\label{eq:q-above-final}
    q[\boldsymbol{m}] = \sum_{\boldsymbol{k}\in \Lambda_+^{\boldsymbol{m}}}\beta[\boldsymbol{k}] \eta[\boldsymbol{k} - \boldsymbol{k}_{\text{T}}],
\end{equation}
where we denote by $\Lambda_+^{\boldsymbol{m}}$ the discrete upper semicircle.
It is the set of pixel indices $\boldsymbol{k}$ whose corresponding pixels intersect with the continuous upper semicircle $C_+^{\mathbf{r}_{\text{T}, \boldsymbol{m}}}$
\begin{equation}
\label{eq:discrete-upper-circle}
     C_+^{\mathbf{r}_{\text{T}, \boldsymbol{m}}}= \left\{ (x, z): (x - x_{\text{T}, m_1})^2 + (z - z_{\text{F}})^2 = r_{\tau}^2 \text{ and } z< z_{\text{F}}\right\},
\end{equation}
where $x_{\text{T}, m_1}$ is the location of the transducer on the measurement grid.

Similarly, for the case of the lower semicircle, we have
\begin{equation}
    \label{eq:q-below-final}
    q[\boldsymbol{m}] = \sum_{\boldsymbol{k}\in \Lambda_-^{\boldsymbol{m}}}\beta[\boldsymbol{k}] \eta[\boldsymbol{k} - \boldsymbol{k}_{\text{T}}],
\end{equation}
where $\Lambda_-^{\boldsymbol{m}}$ is the discrete lower semicircle whose corresponding pixels intersect with the continuous lower semicircle $C_-^{\mathbf{r}_{\text{T}, \boldsymbol{m}}}$
\begin{equation}
\label{eq:discrete-lower-circle}
     C_-^{\mathbf{r}_{\text{T}, \boldsymbol{m}}}= \left\{ (x, z): (x - x_{\text{T}, m_1})^2 + (z - z_{\text{F}})^2 = r_{\tau}^2 \text{ and } z> z_{\text{F}}\right\},
\end{equation}
In our implementation, $\Lambda_+^{\boldsymbol{m}}$ and $\Lambda_-^{\boldsymbol{m}}$ are generated via a function 
from the imaging processing library \emph{skimage} \cite{draw_circle}.
Equation \eqref{eq:q-above-final} (and similarly, \eqref{eq:q-below-final}) can be implemented based on the following steps:
For each measurement location $\boldsymbol{m}$,  
1) Shift $\eta$ to grid location of the transducer, then, element-wise multiply $\beta$ with it;
2) Generate the semicircle mask $\Lambda_+^{\boldsymbol{m}}$ and apply it to the outcome of step 1);
3) Sum up all masked pixels $\boldsymbol{k}$ to obtain $q[\boldsymbol{m}]$.

Since the radius of the semicircle is the same for a given depth, we can generate a row of the image $q$ at once.
To avoid explicitly constructing the forward matrix by computing \eqref{eq:q-above-final} for each pixel $\boldsymbol{m}$, 
we interpret \eqref{eq:q-above-final} equivalently as follows:
the pixel value $q[\boldsymbol{m}]$ is the result of the correlation between a 2D image $\beta$ and a 2D image $\eta$ multiplied by a binary mask, then evaluated at location $\boldsymbol{k}_{\text{T}}$.
We thus implement \eqref{eq:q-above-final} (similarly \eqref{eq:q-below-final}) based on the following steps:
1) Generate the semicircle mask at the current depth $k_{2}$;
2) Element-wise multiply it with the SPR image (see Figure \ref{fig:acoustic-signal-generation} (c) and (c'));
3) Correlate the OA image $\beta$ with the masked SPR image;
4) Extract the horizontal slice at the focal plane $k_{\text{T}, 2}$ from the outcome of Step 2);
5) Perform Step 1) to 4) for all depths and stack the horizontal slices according to their respective location to produce the 2D image $q$.
The temporal convolution step is applied to each column of the 2D image $q$ via standard convolution in 1D.
We provide an illustration of the acoustic signal-generation process in Figure \ref{fig:acoustic-signal-generation}.

We write the forward model for a single acoustic process at switching time point $t_n$ as the linear system
\begin{equation}
    \label{eq:acoustic-discrete-single}
    \mathbf{p}^n = \mathbf{W} \mathbf{b}^n,
\end{equation}
where $\mathbf{b}^n$ and $\mathbf{p}^n \in\mathbb{R}^{L}$ are the vector representation of the OA image $\beta$ and the acoustic image $p$ at $t_n$, respectively.
We construct the acoustic forward operator $\mathbf{W}\in\mathbb{R}^{L \times L}$ as a matrix-free linear operator following the interpretation in Figure \ref{fig:acoustic-signal-generation}.

Finally, we build the acoustic forward operator $\mathbf{W}_{\text{tot}}$ for $N$ switching pulses during an OFF-switching cycle and obtain 
\begin{equation}
    \label{eq:acoustic-full-forward}
    \underbrace{
    \begin{bmatrix}
        \mathbf{p}^{0} \\
        \mathbf{p}^{1} \\
        \vdots \\
        \mathbf{p}^{N-1}
    \end{bmatrix}
    }_{\mathbf{p}}
    =
    \underbrace{
    \begin{bmatrix}
        \mathbf{W} & 
        \mathbf{0} &     
        \cdots & 
        \mathbf{0} \\
        \mathbf{0} & 
        \mathbf{W} &
        \cdots & 
        \vdots \\
        \vdots & 
        \cdots &    
        \ddots &
        \mathbf{0} \\
        \mathbf{0} & 
        \cdots & 
        \mathbf{0} &     
        \mathbf{W} \\
    \end{bmatrix}}
    _{\mathbf{W}_{\text{tot}}}
    \underbrace{
    \begin{bmatrix}
        \mathbf{b}^{0} \\
        \mathbf{b}^{1} \\
        \vdots \\
        \mathbf{b}^{N-1}
    \end{bmatrix}}
    _{\mathbf{b}}  + \quad\mathbf{n}.
\end{equation}
There $\mathbf{b}\in\mathbb{R}^{NL}$ is the vector of the initial pressure of all $N$ pulses, $\mathbf{W}\in\mathbb{R}^{NL\times NL}$ is the system matrix, $\mathbf{p}\in\mathbb{R}^{NL}$ is the vector of the acoustic measurement of all $N$ pulses, and the vector
$\mathbf{n}\in\mathbb{R}^{NL}$ represents measurement noise.
Note that both $\mathbf{W}$ and $\mathbf{W}_{\text{tot}}$ are matrix-free linear operators.

\subsubsection{Complete Forward Pipeline}
We denote the forward operator of the complete pipeline as $\mathbf{A}\in\mathbb{R}^{NL\times PL}$.
It is a composition of the photo-switching forward operator $\mathbf{S}$ and the complete acoustic forward matrix $\mathbf{W}_{\text{tot}}$ written as
\begin{equation}
\label{eq:full-forward}
    \mathbf{A} = \mathbf{W}_{\text{tot}}\mathbf{S}.
\end{equation}
The forward pipeline takes the concentration maps $\mathbf{c}$ as input, applies the photo-switching operator $\mathbf{S}$ in which the optical process is applied $N$ times for $N$ photo-switching pulses to obtain a stack of $N$ OA images $\mathbf{H}$ (or equivalently $\mathbf{b}$, cf. \eqref{eq:pa-effect}).
They are then fed to the acoustic module where the acoustic-detection process is applied to each OA image.
The final measurement is a stack of acoustic signals $\mathbf{p}$.

\subsection{Inverse Problem and Reconstruction Algorithm}

The goal of unmixing is to recover the unknown concentration maps $\mathbf{c}$ from the acoustic measurements $\mathbf{p}$ via the linear system
\begin{equation}
    \label{eq:full-forward-system}
    \mathbf{p} = \mathbf{W}_{\text{tot}}\mathbf{S} \mathbf{c} + \mathbf{n},
\end{equation}
where $\mathbf{n}$ represents measurement noise.

\subsubsection{Two-Step Approach}
In the two-step approach, one states a minimization problem to find the solution $\bar{\mathbf{b}}$ to the acoustic problem in \eqref{eq:acoustic-full-forward}
\begin{equation}
    \label{eq:acoustic-inverse}
    \overline{\mathbf{b}} \in \arg\min_{\mathbf{b}\in\mathbb{R}^{NL}} \left\{\frac{1}{2} \|\mathbf{W}_{\text{tot}} \mathbf{b} - \mathbf{p}\|^2_2 + \mathcal{R}_1(\mathbf{b}) + \delta_{\geq 0}(\mathbf{b}) \right\},
\end{equation}
where the indicator function $\delta_{\geq 0}(\mathbf{x})$ for a vector $\mathbf{x}\in\mathbb{R}^N$is defined as
\begin{equation}
    \delta_{\geq0}(\mathbf{x}) = \left\{
    \begin{aligned}
        &0, \quad \text{if } x_n \geq 0, \quad n=1, \ldots, N\\
        &+\infty, \quad \text{else}.
    \end{aligned}
    \right.
\end{equation}
One then solves the unmixing problem by determining
\begin{equation}
    \label{eq:unmixing}
    \overline{\mathbf{c}} \in \arg\min_{\mathbf{c}\in\mathbb{R}^{PL}} \left\{\frac{1}{2} \|\mathbf{S} \mathbf{c} - \overline{\mathbf{b}}\|^2_2 + \mathcal{R}_2 (\mathbf{c}) + \delta_{\geq 0}(\mathbf{c})\right\}.
\end{equation}
The regularization terms $\mathcal{R}_1(\cdot)$  and $\mathcal{R}_2(\cdot)$ are optional.
Here, we choose to apply total variation (TV) on $\mathbf{b}$ to encourage smoothness in the reconstructed images, and a combination of $l_1$ and TV on $\mathbf{c}$ which was shown effective in \cite{liu_model-based_2025} to improve the quality of reconstruction:
\begin{equation}
    \label{eq:reg-tv}
    \mathcal{R}_1(\mathbf{b}) = \nu_1
    \left(\sum_{n=0}^{N-1}|\mathbf{b}^n|_\text{TV}\right), 
\end{equation}
\begin{equation}
    \label{eq:reg-unmixing}
    \mathcal{R}_2(\mathbf{c}) = \nu_2 
    \left( \sum_{i_1=0}^{L_{\text{x}} - 1} \sum_{i_2=0}^{L_{\text{z}} - 1} \|\mathbf{c}_{i_1, i_2}\|_1 \right)
    + \nu_3 \left(\sum_{p=1}^{P}|\mathbf{c}^p|_\text{TV}\right), 
\end{equation}
where $\nu_1, \nu_2$ and $\nu_3$ are nonnegative regularization weights.
We adopt the anisotropic TV for computational speed. It enforces sparsity in the gradient domain and is defined as
\begin{equation}
    |\cdot|_{\text{TV}} = \|\nabla_\text{x} (\cdot)\|_1 + \|\nabla_\text{z} (\cdot)\|_1,
\end{equation}
where $\nabla_\text{x}$ and $\nabla_\text{z}$ are the finite difference operators in lateral and axial directions, respectively.

\subsubsection{One-Step Approach}
We formulate the outcome of the one-step inversion as the solution $\hat{\mathbf{c}}$ to the minimization problem
\begin{equation}
    \label{eq:one-step}
    \hat{\mathbf{c}} \in \arg\min_{\mathbf{c}\in\mathbb{R}^{PL}} \left\{\frac{1}{2} \|\mathbf{A} \mathbf{c} - \mathbf{p}\|^2_2 + \mathcal{R}_3(\mathbf{c}) +\delta_{\geq0}(\mathbf{c}) \right\},
\end{equation}
where $\mathcal{R}_3(\mathbf{c})$ is a regularization term defined as
\begin{equation}
    \label{eq:reg-tv-l1-one}
    \mathcal{R}_3(\mathbf{c}) = \lambda_1
    \left(\sum_{p=1}^P |\mathbf{c}^{p}|_\text{TV}\right)
    +
    \lambda_2 \left(\sum_{i_1=0}^{L_{\text{x}} - 1} \sum_{i_2=0}^{L_{\text{z}} - 1} \|\mathbf{c}_{i_1, i_2}\|_1
    \right)
    + \lambda_3 \left(\sum_{n=0}^{N-1}|(\mathbf{S}\mathbf{c})^n|_{\text{TV}}\right), 
\end{equation}
and where the nonnegative constants $\lambda_1$, $\lambda_2$, and $\lambda_3$
are the respective regularization weights.
The first term in \eqref{eq:reg-tv-l1-one} applies TV to the spatial concentration maps $\mathbf{c}^p\in\mathbb{R}^{L}$, $p=1, \ldots, P$ of all $P$ species to reduce noise and to achieve a smooth reconstruction.
The second term in \eqref{eq:reg-tv-l1-one} applies the sparsity-promoting $l_1$-norm to each pixel in the concentration map $\mathbf{c}_{i_1, i_2}\in\mathbb{R}^P$, $i_1=0, \ldots, (L_{\text{x}} - 1), i_2=0, \ldots, (L_{\text{z}}-1)$, in order to minimize the cross-talk between species.
The last regularizer in \eqref{eq:reg-tv-l1-one} applies TV to the spatial intensity maps of OA images $\mathbf{H}^n=(\mathbf{S}\mathbf{c})^n\in\mathbb{R}^{L}$, $n=0, \ldots, (N-1)$ of all $N$ switching pulses to enforce a smoothing effect on the intermediate OA images.

\subsubsection{Algorithm}
\begin{algorithm}[t]
\caption{Proximal-gradient algorithm for the main problem \cite{fista}}
\label{alg:prox-outer}
    \begin{algorithmic}[1]
        \State \textbf{Input} initial guess $\mathbf{c}_0=\mathbf{0}$, $\boldsymbol{\xi}_0=\mathbf{0}$, the maximal number $K_1$ of iterations, stopping thresholds $\varepsilon_1>0$ and $\varepsilon_2>0$
        \State \textbf{Set} $k=0$, $t_0=1$, cost $f_0=\infty$
        \State \textbf{Compute} stepsize $\alpha_1=1/\text{eig}_{\max}(\mathbf{A}^T\mathbf{A})$
        \While {$k \le K_1$}:
            \State $\mathbf{c}_{k+1} = \text{prox}_{\alpha_1 w} \left(\ \boldsymbol{\xi}_k - \alpha_1 \left(\mathbf{A}^T\mathbf{A} \boldsymbol{\xi}_k - \mathbf{A}^T \mathbf{p}\right)\right)$
            \State $t_{k+1} = \frac{1+\sqrt{4 t_k^2 + 1}}{2}$
            \State $\boldsymbol{\xi}_k = \mathbf{c}_{k+1} + \frac{t_k - 1}{t_k + 1}(\mathbf{c}_{k+1} - \mathbf{c}_k)$
            \If {$k>1$}
                \If {$\frac{|f_k - f_{k-1}|}{|f_{k-1}|} < \varepsilon_1$ \textbf{or} $\frac{\|\boldsymbol{\xi}_{k} - \boldsymbol{\xi}_{k-1}\|_2}{\|\boldsymbol{\xi_{k-1}}\|_2} < \varepsilon_2$}
                \State \textbf{break}
                \EndIf 
            \EndIf
            \State $k\leftarrow k+1$
        \EndWhile
        \State \textbf{Output} $\mathbf{c}_{k+1}$
    \end{algorithmic}
\end{algorithm}

\begin{algorithm}[t]
\caption{Computation of the proximal operator \cite{tibshirani_solution_2011}}
\label{alg:prox-inner}
    \begin{algorithmic}[1]
        \State \textbf{Input} initial guess $\mathbf{u}_0=\mathbf{0}$, $\boldsymbol{\eta}_0=\mathbf{0}$, and 
        the maximal number $K_2$ of iterations
        \State \textbf{Set} $t_0=1$
        \State \textbf{Compute} stepsize $\alpha_2=1/\text{eig}_{\max}(\mathbf{L}^T\mathbf{L})$
        \For {$k=0$ to $K_2$}:
            \State $\mathbf{u}_{k+1} = \text{proj}_{\alpha_1} \left(\ \boldsymbol{\eta}_k - \alpha_2 \left(\mathbf{L}\mathbf{L}^T \boldsymbol{\eta}_k - \mathbf{L} \mathbf{z}\right)\right)$
            \State $t_{k+1} = \frac{1+\sqrt{4 t_k^2 + 1}}{2}$
            \State $\boldsymbol{\eta}_k = \mathbf{u}_{k+1} + \frac{t_k - 1}{t_k + 1}(\mathbf{u}_{k+1} - \mathbf{u}_k)$            
        \EndFor
        \State \textbf{Output} $\max(\mathbf{z} - \mathbf{L}^T \mathbf{u}_{k+1}, \mathbf{0})$
    \end{algorithmic}
\end{algorithm}
The objective functions of the minimization problems in \eqref{eq:acoustic-inverse}, \eqref{eq:unmixing}, and \eqref{eq:one-step} share a similar structure of a smooth part (the quadratic data-fidelity term), which we denote by $f$, and a nonsmooth part (the sparsity-based regularization term and the nonnegativity constraint), which we denote by $w$.
Further, the three regularizers introduced in \eqref{eq:reg-tv},
 \eqref{eq:reg-unmixing}, and \eqref{eq:reg-tv-l1-one}
can be rewritten in the form of the $L_1$-norm of an operator $\mathbf{L}$ as $\mathcal{R}(\cdot) = \|\mathbf{L}(\cdot)\|_1$.
For example, $\mathbf{L}$ for $\mathcal{R}_3$ is
\begin{equation}
    \mathbf{L} = \begin{bmatrix}
        \mathbf{L}_1 \\
        \mathbf{L}_2 \\
        \mathbf{L}_3
    \end{bmatrix},
\end{equation}
where
\begin{equation}
    \mathbf{L}_1 = \lambda_1
    \begin{bmatrix}
        \nabla_\text{x}\\
        \nabla_\text{z}
    \end{bmatrix}, \quad
    \mathbf{L}_2 = \lambda_2
    \begin{bmatrix}
        \nabla_\text{x} \cdot \mathbf{S}\\
        \nabla_\text{z} \cdot \mathbf{S}
    \end{bmatrix}, \quad
    \mathbf{L}_3 = \lambda_3 \mathbf{I},
\end{equation}
and $\mathbf{I}$ is the identity operator.
The operators $\mathbf{L}$ of $\mathcal{R}_1$ and $\mathcal{R}_2$ are similar and thus omitted here in the interest of space.

We use \eqref{eq:one-step} as an example to show how to obtain the solution, as \eqref{eq:acoustic-inverse} and \eqref{eq:unmixing} follow the same approach.
The objective function in \eqref{eq:one-step} is composed of a smooth part $f(\cdot)=\frac{1}{2} \|\mathbf{A} \cdot - \mathbf{p}\|^2_2$ and a nonsmooth part $w(\cdot)=\mathcal{R}_3(\cdot) + \delta_{\geq0}(\cdot)$.
We thus deploy a proximal-gradient method combined with the fast iterative shrinkage thresholding algorithm (FISTA) \cite{fista} to obtain the solution.
Detailed steps are presented in Algorithm \ref{alg:prox-outer}.
There, the key is the computation of the proximal operator $\text{prox}_{\alpha_1 w}(\mathbf{\cdot})$ which is defined as
\begin{equation}
\label{eq:prox}
    \text{prox}_{\alpha_1 w}(\mathbf{z}) = 
    \arg\min_{\mathbf{y}\in\mathbb{R}^{PL}}\left\{
    \frac{1}{2} \|\mathbf{y} - \mathbf{z}\|_2^2 + \alpha_1 \mathcal{R}_3(\mathbf{y}) + \delta_{\geq0}(\mathbf{y})
    \right\},
\end{equation}
where $\alpha_1>0$ is the stepsize and we set it as the reciprocal of the largest eigenvalue of $\mathbf{A}^T\mathbf{A}$. 
To obtain the proximal operator, we resort to the dual problem of the minimization in \eqref{eq:prox} instead \cite{tibshirani_solution_2011} and establish the solution to the dual problem of \eqref{eq:prox} as
\begin{equation}
\label{eq:prox-dual}
    \bar{\mathbf{u}} \in \text{arg}\min_{\mathbf{u}\in\mathbb{R}^{2NL+3PL}} \left\{\frac{1}{2}\| \mathbf{z} - \mathbf{L}^T \mathbf{u} \|^2_2\right\}, \quad \text{s.t.} \quad \|\mathbf{u}\|_{\infty} \le \alpha_1 \text{ and } \mathbf{L}^T\mathbf{u}<\mathbf{z}.
\end{equation}
We solve the dual problem using accelerated gradient descent (another FISTA, similar to the approach in \cite{pourya_delaunay_2023}) followed by a projection onto the $L_{\infty}$-ball during optimization,
where the projection operator $\text{proj}_{\alpha}(\cdot)$ applies element-wise to a vector $\mathbf{x}\in\mathbb{R}^N$ and is defined as
\begin{equation}
    \text{proj}_{\alpha}(\mathbf{x})_n = \left\{
    \begin{aligned}
        & x_n, \quad |x_n| < \alpha \\
        & \alpha, \quad x_n > \alpha \\
        & -\alpha, \quad x_n < -\alpha
    \end{aligned}
    \right.
    \quad n=1, \ldots, N.
\end{equation}
Finally, we retrieve the proximal operator in \eqref{eq:prox} via $\text{prox}_{\alpha w}(\mathbf{z})=\max(\mathbf{z} - \mathbf{L}^T\bar{\mathbf{u}}, \mathbf{0})$ (See Algorithm \ref{alg:prox-inner}).

\section{Results}
\subsection{Setup}
\begin{figure}[t!]
    \centering
    \includegraphics[width=\linewidth]{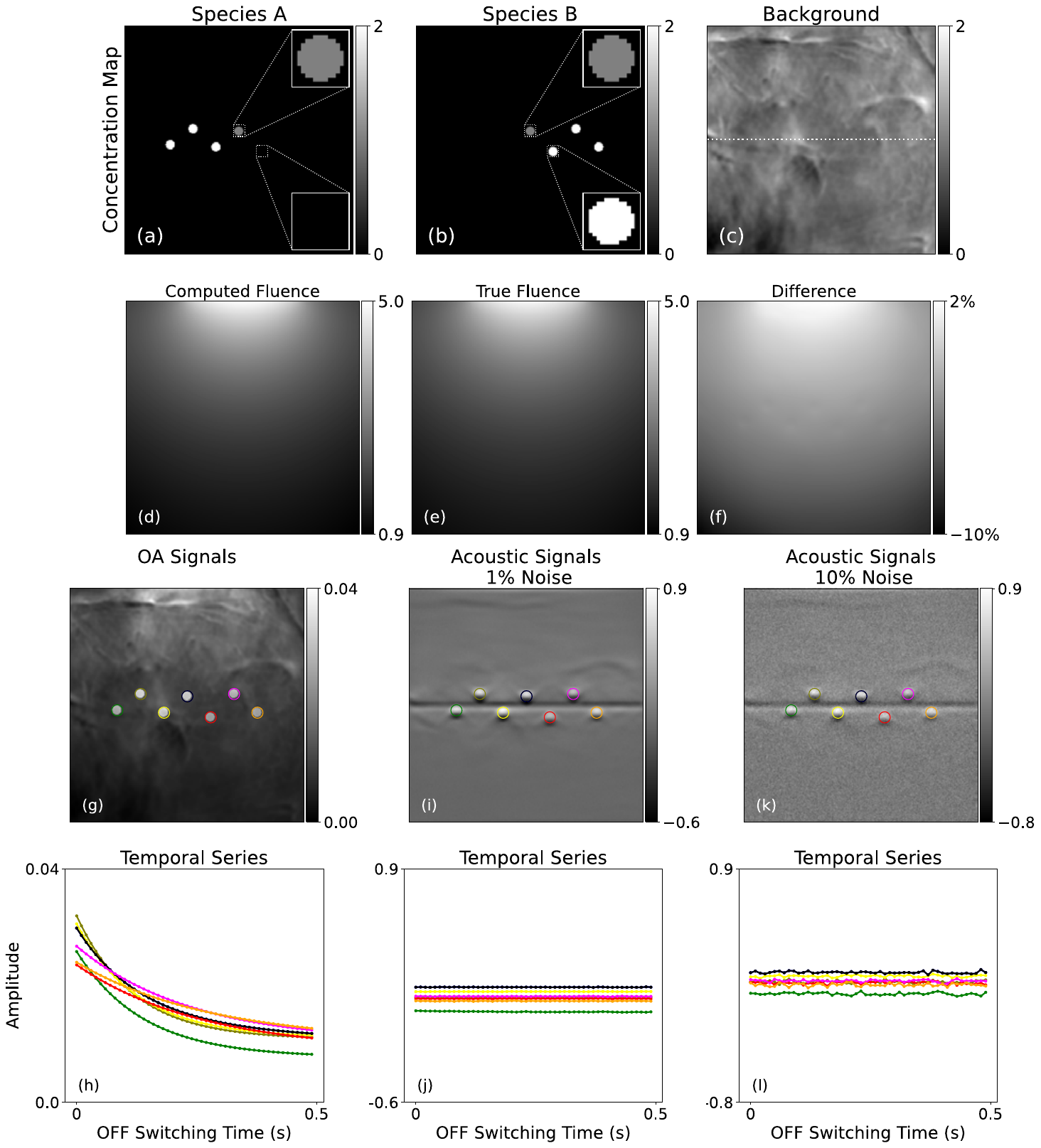}
    \caption{
    (a)-(c) Ground-truth concentration maps of the two photo-switching species of reporters and the non-switching background. The unit is $\mu$M (micromolar).
    The insets of two of the reporters are shown in (a) and (b) for better visualization.
    The horizontal dotted line in (c) indicates the position of the focal plane of the transducer.
    (d) Computed fluence map (arbitrary unit) used for reconstruction.
    (e) True fluence map (arbitrary unit) used to synthesize measurements.
    (f) Difference map between (d) and (e) in percentage.
    (g) Synthesized OA signals (first frame) during an OFF-switching cycle.
    Each reporter is circled out for better identification.
    (h) Temporal series of the OA signals (intensity averaged over the area of each reporter).
    The color coding is the same as in (g), similar for (j) and (l).
    (i) and (k): Subsequent acoustic signals (first frame) with 1\% (i) and 10\% (k) noise.
    (j) and (l): Temporal series of the acoustic signals (intensity averaged over the area of each reporter) that correspond to (i) and (k).
    }
    \label{fig:setup}
\end{figure}

We use a numerical phantom to represent a 2D sample with physical size $(5.6 \times 5.6)$ mm\textsuperscript{2} (numerical size $(300 \times 300)$ pixels) and disk-like objects to mimic photo-switching reporters.
They are located around the acoustic focal plane, at depth 2.8 mm.
We also consider two species A and B.
They are located in the disks on top of a heterogeneous background,
One disk in particular contains a mixture of A and B in 1:1 ratio, which helps us to test the performance of the algorithm on spatially overlapping targets (see Figure \ref{fig:setup} (a)-(c)).
We let that species A has a higher dynamic range of switching signals and a faster switching speed than species B.
Detailed information on the photo-physical properties and optical parameters involved in the simulation is provided in Table \ref{tab:photophysics} in Appendix C.
We use a uniform illumination of width $W=2.8$ mm centered on the top surface of the sample.
The computed fluence map, assuming constant absorption coefficient $\mu_{\text{a}}=0.02$ mm\textsuperscript{-1} and constant scattering coefficient $\mu_{\text{s}}=1$ mm\textsuperscript{-1} maps, is used for reconstruction and shown in Figure \ref{fig:setup} (d).
We compare it with the true fluence map (Figure \ref{fig:setup} (e)) computed using the true heterogeneous absorption coefficient map based on the the background and the reporters and show the difference between them in Figure \ref{fig:setup} (f).
We will see that despite of the mismatch of up to 10\% between the approximated and the true fluence distribution, our framework is robust to achieve good unmixing results.

To synthesize the acoustic measurements for the reconstruction, we construct and apply the true forward model on the ground-truth concentration maps shown in Figure \ref{fig:setup} (a)-(c) using the true fluence distribution map.
Gaussian random noise proportional to a fraction of the maximal amplitude of the complete switching cycle is added to the measurements.
The temporal evolution of photo-switching OA and resulting acoustic signals with 1\% and 10\% noise are shown in Figure \ref{fig:setup} (g)-(l).
Animations of these signals are also available online 
\cite{code}.
The forward model for the reconstruction algorithm is established using the computed fluence map without prior information on the heterogeneity of the background or the reporters.

We use four metrics to evaluate the quality of the reconstruction of each species compared with the ground truth.
This affords us several perspectives.
\begin{itemize}
    \item The normalized root-mean square error (NRMSE) quantifies the relative total error of the reconstruction.
    \item The peak signal-to-noise ratio (PSNR) evaluates the strength of the signal against noise.
    \item The structured-similarity index (SSIM) measures the textural similarity between the reconstruction and the ground truth.
    \item The Dice similarity (Dice) assesses the overlap of the locations of the signal between the reconstruction and the ground truth. It is between 0 and 1, and a high Dice value indicates good recovery of the location of the region of interest.
\end{itemize}
Detailed definition of the metrics is provided in Appendix B.

\subsection{Reconstruction Results}
\begin{figure}[t!]
    \centering
    \includegraphics[width=0.75\linewidth]{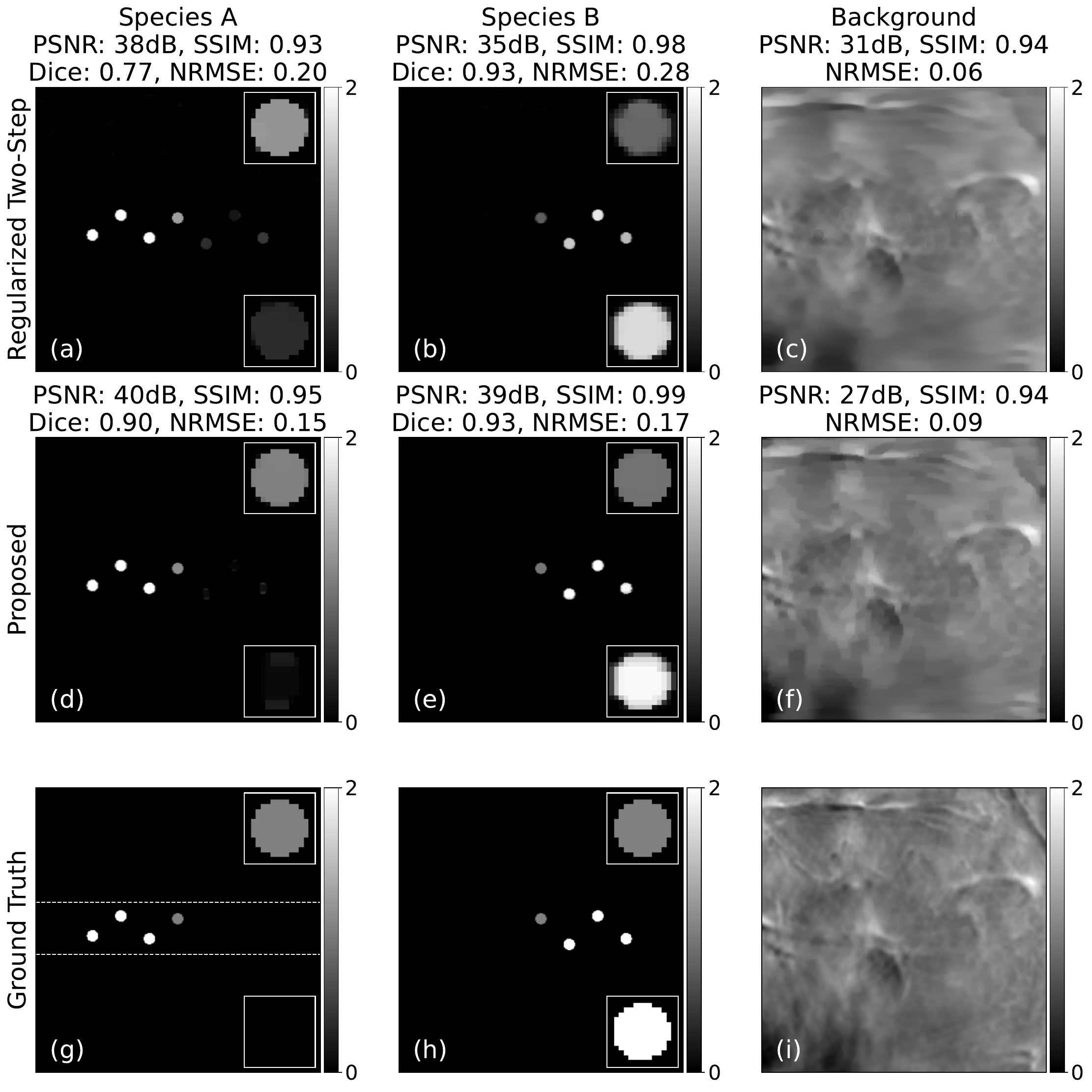}
    \caption{
    Reconstructed concentration maps using the regularized two-step ((a)-(c)) and one-step ((d)-(f))approaches under \textbf{1\% noise} level.
    (g)-(i) Ground truth.
    The rectangular region between the two horizontal dashed lines in (g) indicates the area on which we calculate the SSIM.
    }
    \label{fig:low-noise-recon}
\end{figure}

\begin{figure}[t!]
    \centering
    \includegraphics[width=\linewidth]{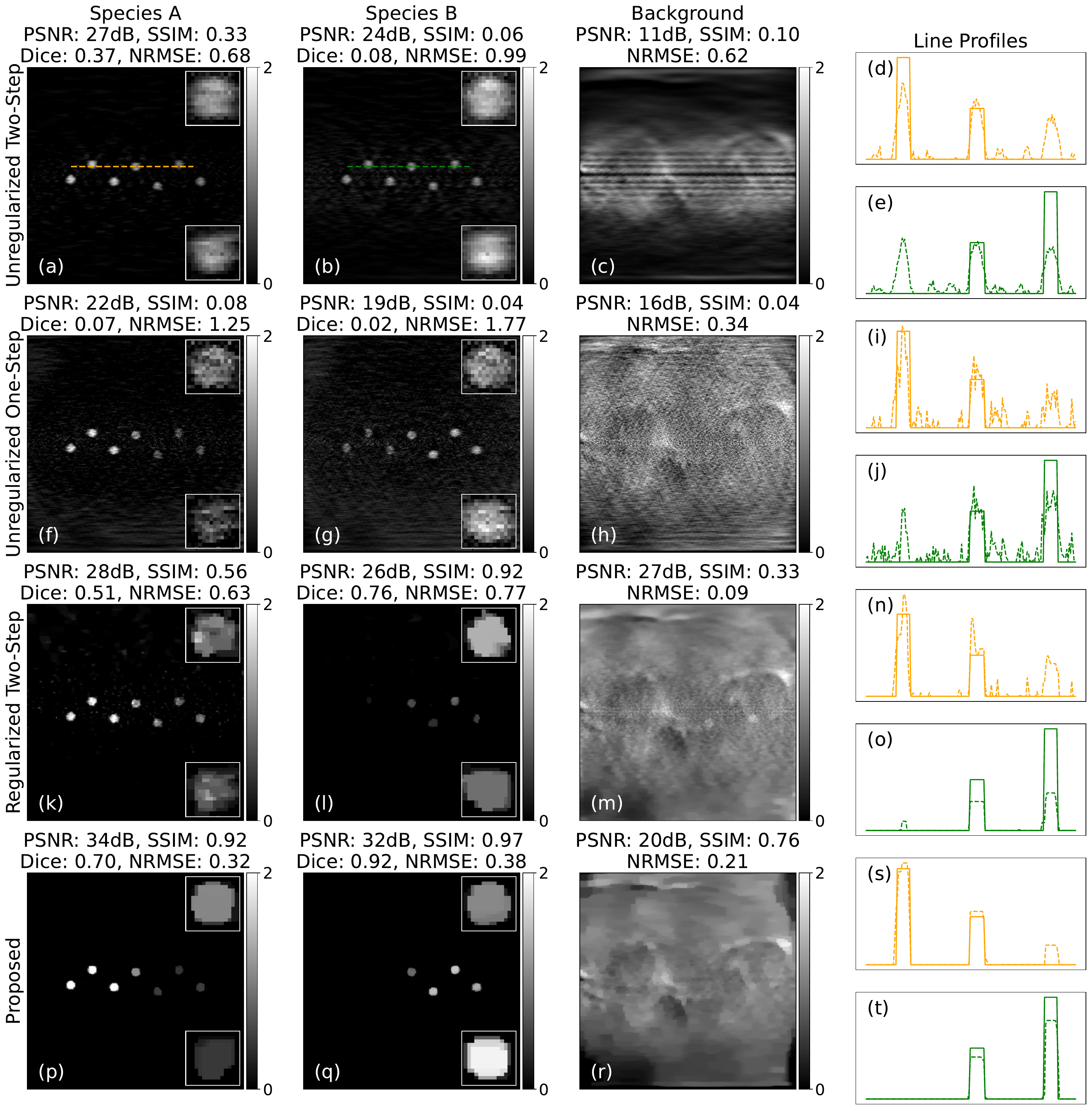}
    \caption{Comparison of the reconstruction results using different inversion approaches under \textbf{10\% noise} level.
    (a)-(e): unregularized two-step approach.
    (f)-(j): unregularized one-step approach.
    (k)-(o): Regularized two-step approach.
    (p)-(t): Regularized one-step approach.
    First to third columns: reconstructed concentration maps of the two species and the background.
    Last column: Intensity of the reconstruction along a dashed line segment drawn in (a) and (b).
    The color orange and green represents the line profiles in species A and B, respectively.
    The corresponding solid orange and green lines represent the ground truth.
    }
    \label{fig:high-noise-recon}
\end{figure}

\subsubsection{Low-Noise Regime}\label{sec:low-noise}
We first show the performance of our proposed regularized one-step approach, as compared to the regularized two-step approach, with 1\% noise in the acoustic signals.
In Figure \ref{fig:low-noise-recon}, we see that our method not only faithfully recovered the concentration maps of the two photo-switching species compared to the ground truth, but also cleanly unmixed them from each other and from the background.
Our method also performs better than the regularized two-step method in the reconstruction of the two species. 

\subsubsection{High-Noise Regime}\label{sec:high-noise}
We then further validate our approach by comparing it with the two-step approach at a higher level of noise of 10\%.
In Figure \ref{fig:high-noise-recon}, we compare four inversion techniques: unregularized two-step, unregularized one-step, regularized two-step, and regularized one-step.
The unregularized techniques use the LSQR algorithm \cite{paige_lsqr_1982} to obtain the solution.
We observe that only our proposed regularized one-step approach successfully reconstructed and unmixed the two species out of the background (Figure \ref{fig:high-noise-recon} (p)-(t)); a close comparison of the intensity profile over a line segment (indicated in Figure \ref{fig:high-noise-recon} (a) and (b)) with the ground truth reveals the quality of reconstruction.
The two unregularized approaches separated the photo-switching reporters from the background but failed to distinguish the two species and mitigate noise in the reconstruction.
The regularized two-step approach produced less noisy images due to the regularization but failed at unmixing the slower-switching species B from the faster-switching species A or from the background.
This is indicated by the absence of reporters that belong to species B in Figure \ref{fig:high-noise-recon} (l) and the appearance of them in image of species A and the background (Figure \ref{fig:high-noise-recon} (k) and (m)).

The evolution of the cost during the main minimization problem for the two regularized methods in Figure \ref{fig:low-noise-recon} and \ref{fig:high-noise-recon} is shown in Figure \ref{fig:cost-evolution} in Appendix C.
The hyperparameters used in the experiments are provided in Appendix D.

\subsubsection{Dependence on the Setup}
\begin{figure}[t!]
    \centering
    \includegraphics[width=\linewidth]{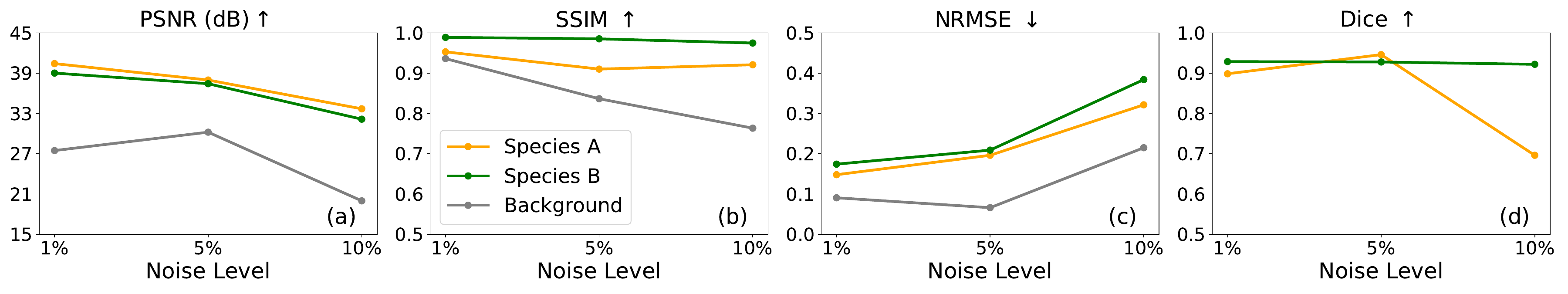}
    \includegraphics[width=\linewidth]{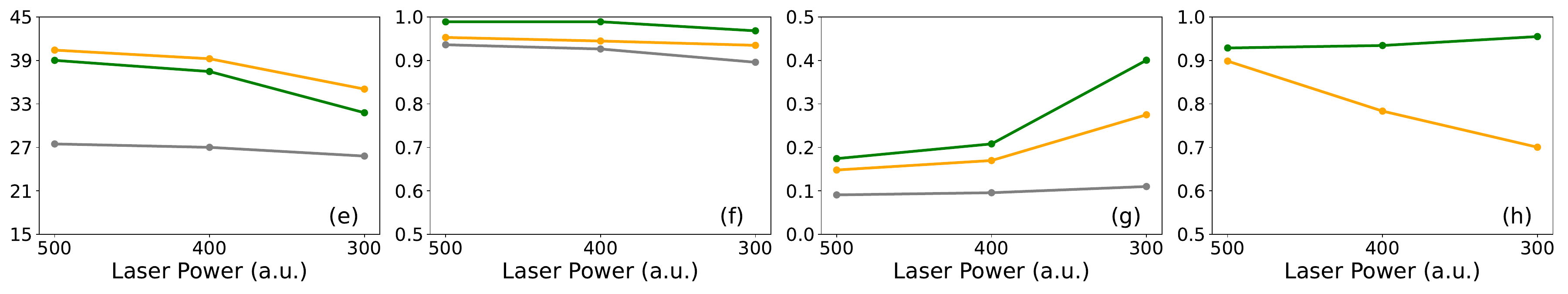}
    \includegraphics[width=\linewidth]{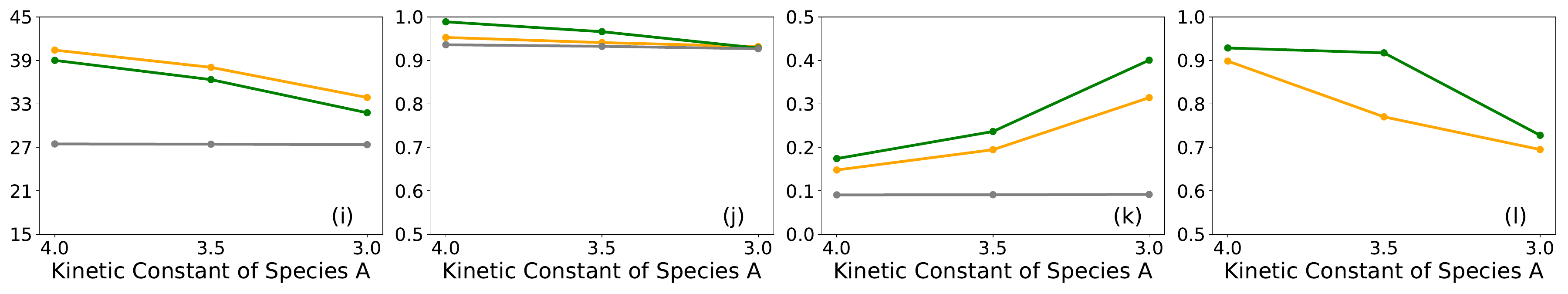}
    \caption{
    Performance of the proposed regularized one-step approach in terms of:
    (a)-(d) noise level;
    (e)-(h) laser power (arbitrary unit);
    (i)-(l) kinetic constant.
    }
    \label{fig:compare-metrics-over-noise-levels}
\end{figure}
Finally, we check the performance of our proposed approach in a variety of configurations.
\begin{itemize}
    \item Noise level: 1\%, 5\%, and 10\%.
    \item Kinetic constant of species A: 4.0, 3.5 and 3.0 (while species B is 2.0).
    \item Laser energy: 500, 400, and 300 (arbitrary unit).
\end{itemize}
We show the quantitative evaluation of our approach for several configurations of these settings in Figure \ref{fig:compare-metrics-over-noise-levels}.
We see that the performance of our approach is stable across the configurations.
In the test of robustness against the noise level, the SSIM and the Dice values are close to the perfect value 1.0 (see Figure \ref{fig:compare-metrics-over-noise-levels} (b) and (d)).
The PSNR values experience a slight decrease when the noise level is raised (Figure \ref{fig:compare-metrics-over-noise-levels} (a)), and the NRMSE increases (Figure \ref{fig:compare-metrics-over-noise-levels} (c)).

In a second test, we show the impact of the laser power.
Similar to the results in the first test, the performance of our approach is still stable, especially in terms of SSIM (Figure \ref{fig:compare-metrics-over-noise-levels} (f)).
When it decreases, the intensity of the light fluence is lowered, which decreases not only the amplitude of the detected signal but also the switching speed.
The difficulty of unmixing therefore increases, which explains the decrease in Dice for species A in Figure \ref{fig:compare-metrics-over-noise-levels} (h).

We also compare the quality of unmixing when the difference between the kinetic constants of the two species changes.
The evolution curve of the three metrics of the background in Figure \ref{fig:compare-metrics-over-noise-levels} (i)-(k) remains more or less flat, which indicates that the reconstruction of the background is not influenced.
The reconstruction quality of the two species, measured by the four metrics, shows similar tendencies as compared with previous tests.
Overall, the performance of our proposed approach is stable across different configurations.

\subsection{Computational Analysis}

\begin{figure}[t]
    \centering
    \includegraphics[width=0.48\linewidth]{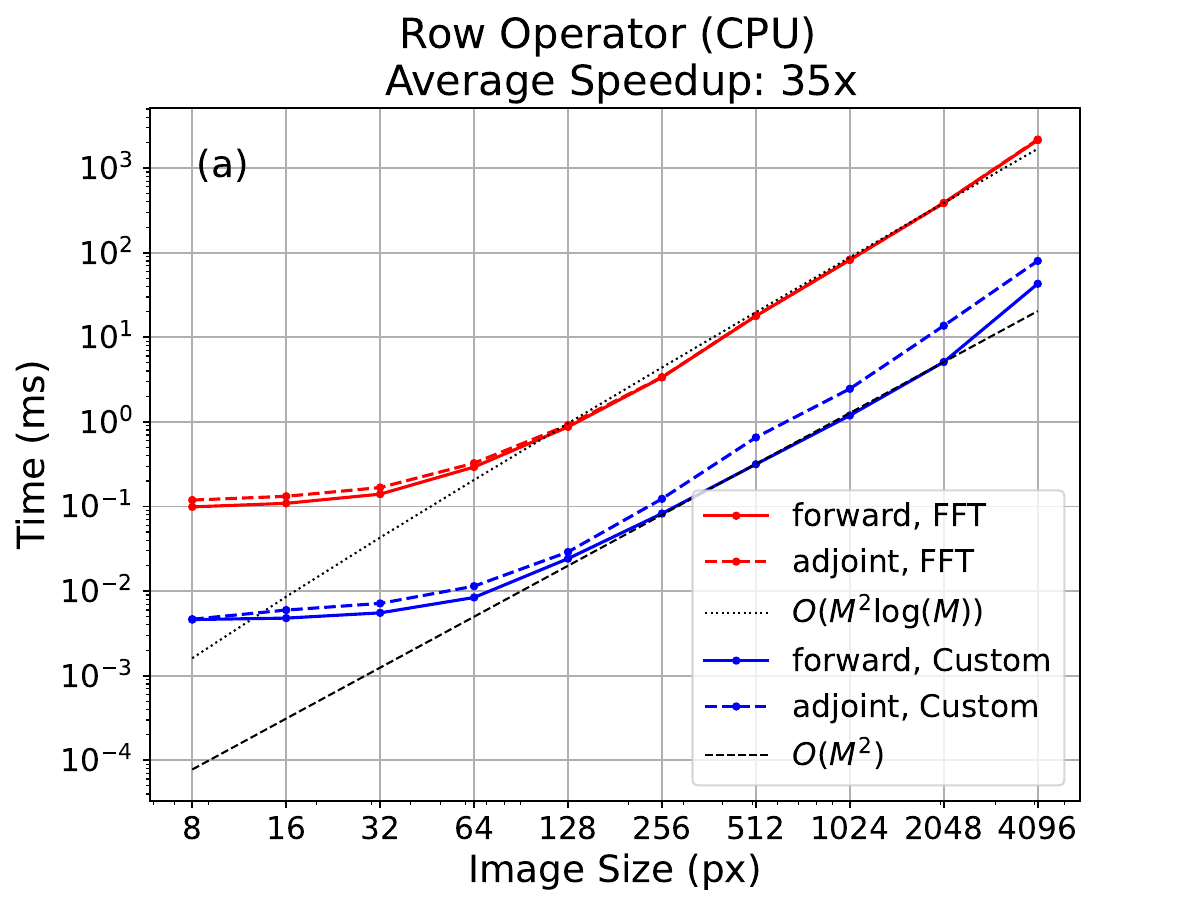}
    \includegraphics[width=0.48\linewidth]{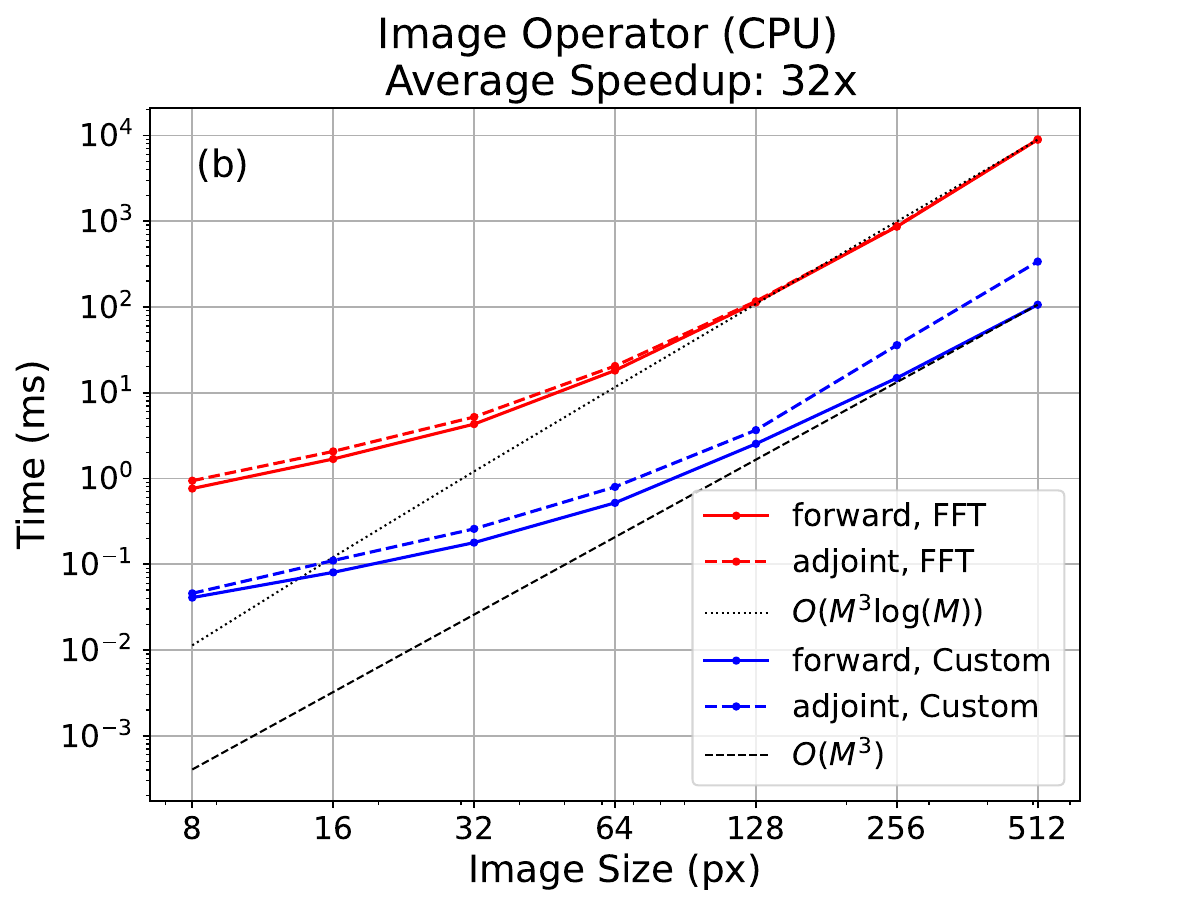}
    \includegraphics[width=0.48\linewidth]{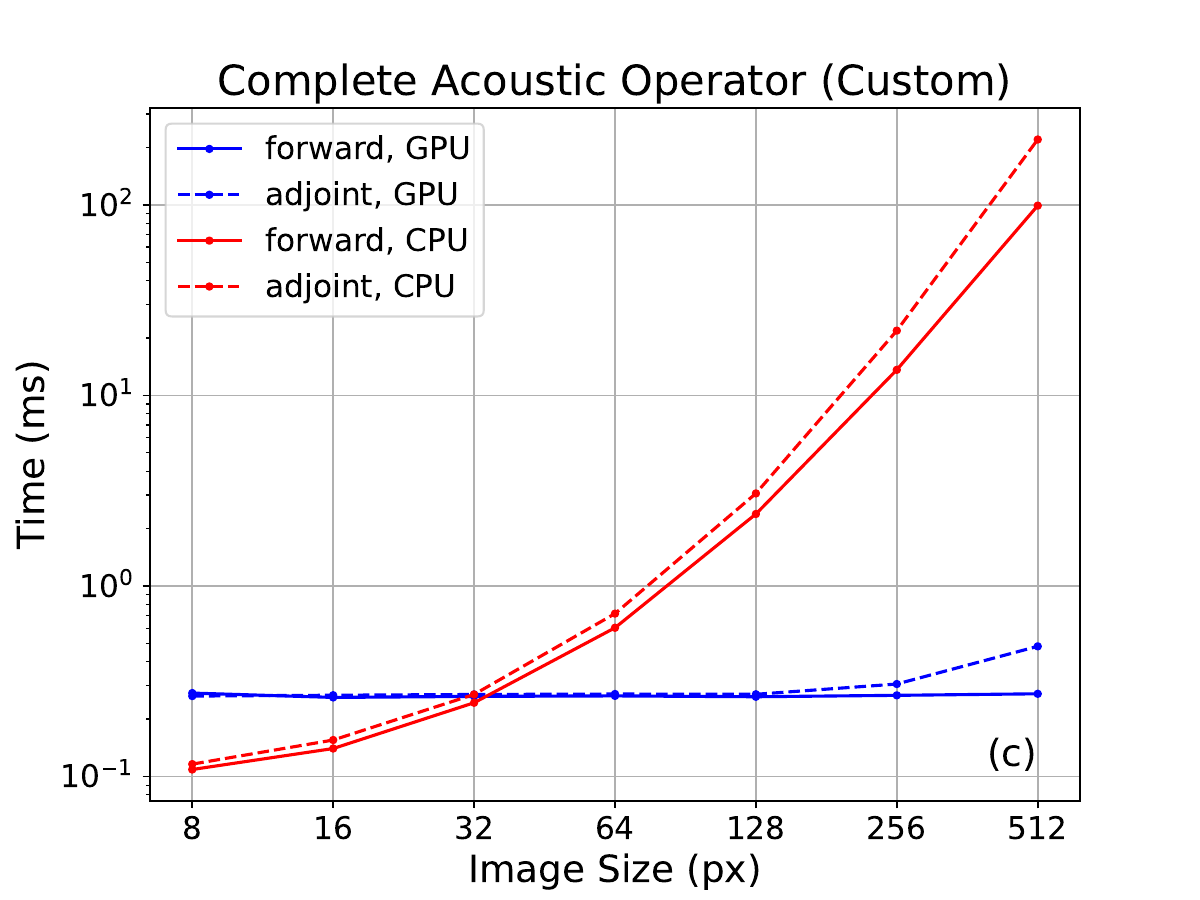}
    \includegraphics[width=0.48\linewidth]{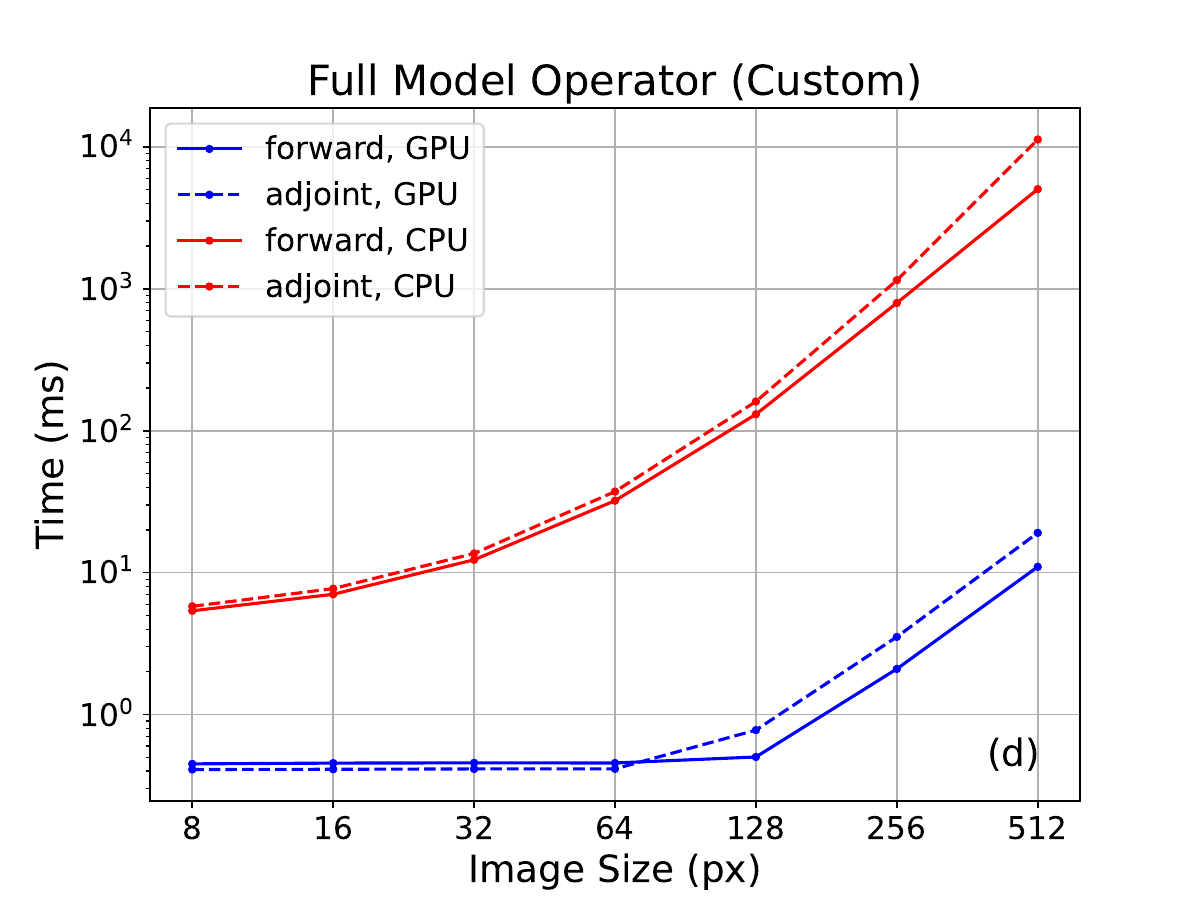}
    \caption{Runtime benchmark of the forward and adjoint modes of the Row Operator (a), the Image Operator (b), the complete acoustic operator (c) and the full-model operator (d).
    The image size refers to the width in pixels of a square image.
    The number $N$ of sampling points in the switching temporal domain is 50 and the number $P$ of species is 3.
    Each data point in these plots represents a value averaged over 10 repetitions.}
    \label{fig:benchmark-rowop}
\end{figure}

For a phantom of size $(300 \times 300)$ pixels and an SPR of size $(300 \times 150)$ pixels, the average computational time for the presented regularized one-step and two-step results is 148 seconds and 117 seconds, respectively.
The spatial integration step of the acoustic model (Equation \eqref{eq:q-above-final} and \eqref{eq:q-below-final}) accounts for the majority of the computations and requires an efficient implementation.
It involves repeated correlations between the input image and a sparse kernel composed of 1D semicircles of various radii that are depth-dependent (see Figure \ref{fig:acoustic-signal-generation}).
The implementation of the spatial integration consists of two core operators: (i) the Row Operator that generates one horizontal slice of the output signal at a given depth via a correlation with the masked spatial response, followed by the extraction of a row of pixels at the focal plane; (ii) the Image Operator assembles the output of the Row Operators at all depths to form a complete 2D image.
One approach to compute the correlation involved in the Row Operator is to use the fast Fourier transform (FFT) and dense array representations of 2D images, hereafter referred to as the FFT Row Operator.
Another approach, referred to as the custom Row Operator, computes correlations directly in the original domain.

Let $M$ be the numerical size of the width of a square image. Then, the theoretical complexities of the FFT Row Operator and Image operator are $O(M^2\log M)$ and $O(M^3\log M)$, respectively, and $O(M^2)$ and $O(M^3)$, respectively, for the custom Row Operator (details are provided Appendix E).
Although the custom approach tends to scale unfavorably in comparison with FFT, especially for kernels of large sizes, our setting permits two important simplifications that can mitigate these effects.
Firstly, we only evaluate the correlated signal on the small window of the output image where it is actually needed, namely, along the row of pixels at the focal plane.
Secondly, we exploit the sparsity of the kernels by storing and processing only their nonzero entries.
Overall, the custom approach achieves better asymptotic scaling and memory efficiency than the FFT one, while also being trivial to parallelize over the image rows.
We use PyLops, an open-source library for the modeling and solving of large-scale linear problems \cite{pylops}, to define the forward models in the complete pipeline as matrix-free operators --- the explicit assembly of the matrices is prohibitive due to their sizes.
We use the open-source library Numba \cite{numba} to achieve the just-in-time compilation to accelerate the custom Row Operator.
We further accelerate it on GPU using CuPy \cite{cupy_learningsys2017} and a handwritten CUDA kernel that performs all the per-row correlations simultaneously and in parallel as the Image Operator.

We benchmarked the forward and adjoint modes of the two versions of the Row Operator and the corresponding Image Operator on an Intel i9-10900X CPU and then the complete acoustic operator and the full pipeline on an NVIDIA GeForce RTX3090 GPU.
The results are shown in Figure \ref{fig:benchmark-rowop}.
We see that they agree well with the theoretical complexity.
The custom Row and Image Operators are approximately 50x faster than the FFT counterparts.
One evaluation of either the forward or the adjoint mode of the custom Row and Image Operator of a sample of $(512\times512)$ pixels with 3 species and 50 switching points costs 0.05 ms and 100 ms, respectively, compared to 10 ms and $10^4$ ms in the case of the FFT approach.
In Figure \ref{fig:benchmark-rowop} (c) and (d), we compare the complete acoustic operator including the temporal response (c) and the complete forward model including the acoustic and the photo-switching modules (d) when using the custom Image operator. 
We see a significant speedup on GPU compared to CPU.
The code to produce the results in the paper is provided in \cite{code}.

\section{Conclusion}
We have presented a comprehensive model for photo-switching optoacoustic mesoscopy and a global inversion framework to reconstruct the concentration maps directly from the acoustic measurements.
Our global framework includes a one-step reconstruction algorithm with a tailored $l_1$ regularization combined with TV regularization on two spaces to mitigate noise and improve the quality of unmixing.
We have shown that our regularized one-step approach is consistently robust as compared to other approaches and under various setups.
In particular, it is also robust to mismatches in the fluence estimation, which is beneficial for realistic experiments.
We provide an efficient GPU implementation of the pipeline.
Its benchmarking results underline its relevance to fast iterative-reconstruction algorithms.
Our framework is extendable to 3D imaging and flexible enough to include other models of the transducer impulse response.
It provides a unique opportunity for in-depth imaging at cellular resolution using photo-switching optoacoustic mesoscopy.

\section{Acknowledgments}
We would like to acknowledge the funding supported by European Union's Horizon Europe Research and Innovation Programme under Grant Agreement No. (101046667 (SWOPT)).
We would like to express our gratitude to our collaborators from Helmholtz Munich, in particular, Prof. André C. Stiel and Dr. Hailong He for fruitful discussions, and Prof. Dominik Justel for providing the data of the spatial response of the transducer.
We also appreciate the kind help in software engineering and manuscript reviewing from Eric Sinner.

\bibliography{biblio.bib}
\bibliographystyle{IEEEtran}

\newpage

\section*{Appendix A}\label{sec:fluence-computation}
We provide in Algorithm \ref{alg:fenics} the details to compute the fluence.
\begin{algorithm}[h]
\caption{Algorithm to solve the diffusion equation using Fenicsx \cite{AlnaesEtal2015}}
\label{alg:fenics}
    \begin{algorithmic}[1]
        \State \textbf{Input} $\mu_{\text{a}}(\mathbf{r}), \mu_{\text{s}}^{'}(\mathbf{r})$, and $S(\mathbf{r})$ as arrays
        \State Define mesh and function space $V$ of type ``continuous Galerkin'' of order 1
        \State Convert $\mu_{\text{a}}(\mathbf{r}), \mu_{\text{s}}^{'}(\mathbf{r})$, and $S(\mathbf{r})$ to functions in $V$
        \State Assemble \eqref{eq:final-pde} into a linear form $a(\Phi, v) = L(v)$
        \State Solve the linear form to get solution $\phi_h$
        \State Convert the finite-element solution $\phi_h$ to an array $\phi$
        \State \textbf{Output} $\phi$
    \end{algorithmic}
\end{algorithm}

\section*{Appendix B}
This section contains the definition of the quantification metrics.
Let vectors $\mathbf{x}, \mathbf{y} \in\mathbb{R}^N$ be the ground truth and reconstructed images.
The NRMSE \cite{harrison_least-squares_2013} is defined as 
\begin{equation}
    \label{eq:nrmse}
    \text{NRMSE} = \frac{\|\mathbf{x} - \mathbf{y}\|_2}{\|\mathbf{x}\|_2}.
\end{equation}
The PSNR is defined as
\begin{equation}
\label{eq:psnr}
    \text{PSNR}(\mathbf{x}, \mathbf{y}) = 20 \cdot \log_{10} (\text{MAX}) - 10 \cdot \log_{10} (\|\mathbf{x} - \mathbf{y}\|_2),
\end{equation}
where MAX represents the largest possible pixel value of the image.
SSIM \cite{zhou_wang_image_2004} is defined as
\begin{equation}
\label{eq:ssim}
    {\rm SSIM}({\bf x},{\bf y})=\frac{(2\mu_{\mathbf{x}}\mu_{\mathbf{y}}+C_{1})(2\sigma_{\mathbf{xy}}+C_{2})}
    {\left(\mu_{\mathbf{x}}^{2}+\mu_{\mathbf{y}}^{2}+C_{1}\right) \left(\sigma_{\mathbf{x}}^{2}+\sigma_{\mathbf{y}}^{2}+C_{2}\right)},
\end{equation}
where $\mu_{\mathbf{x}(\mathbf{y})}$ and $\sigma_{\mathbf{x}(\mathbf{y})}$ are the estimated mean intensity and standard deviation of an image $\mathbf{x}(\mathbf{y})$, $C_1$ and $C_2$ are positive constants to avoid division by too small numbers.
Dice \cite{dice_measures_1945} is defined as 
\begin{equation}
    \label{eq:dice}
      \textrm{Dice}=\frac{2|\textrm{ROI}(\mathbf{y})\cap\textrm{ROI}(\mathbf{x})|}{|\textrm{ROI}(\mathbf{y})|+|\textrm{ROI}(\mathbf{x})|}, \quad \text{ROI}(\mathbf{x}) = \left\{ n : x_n > \max(\mathbf{z}) \times 10\% \right\}.
\end{equation}
Here, we choose a strict threshold of 10\% in the definition of the region of interest (ROI) to better evaluate the separation of the signals in our simulations. 

\section*{Appendix C}
\begin{figure}[h]
    \centering
    \includegraphics[width=0.32\linewidth]{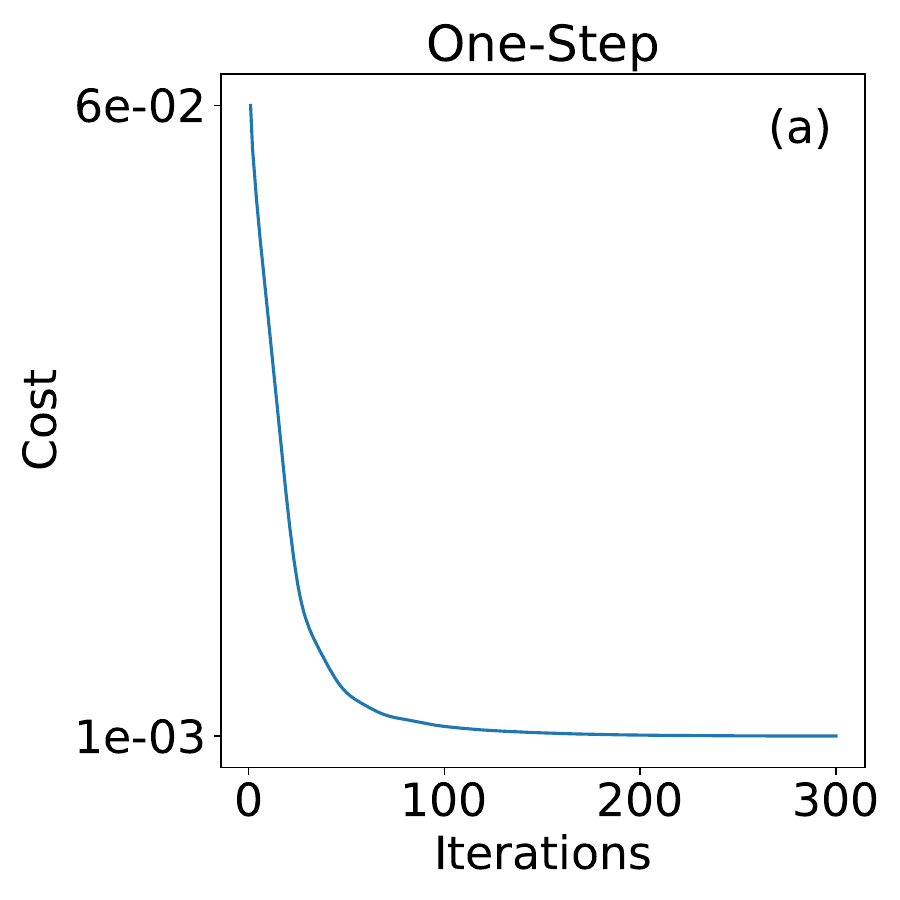}
    \includegraphics[width=0.32\linewidth]{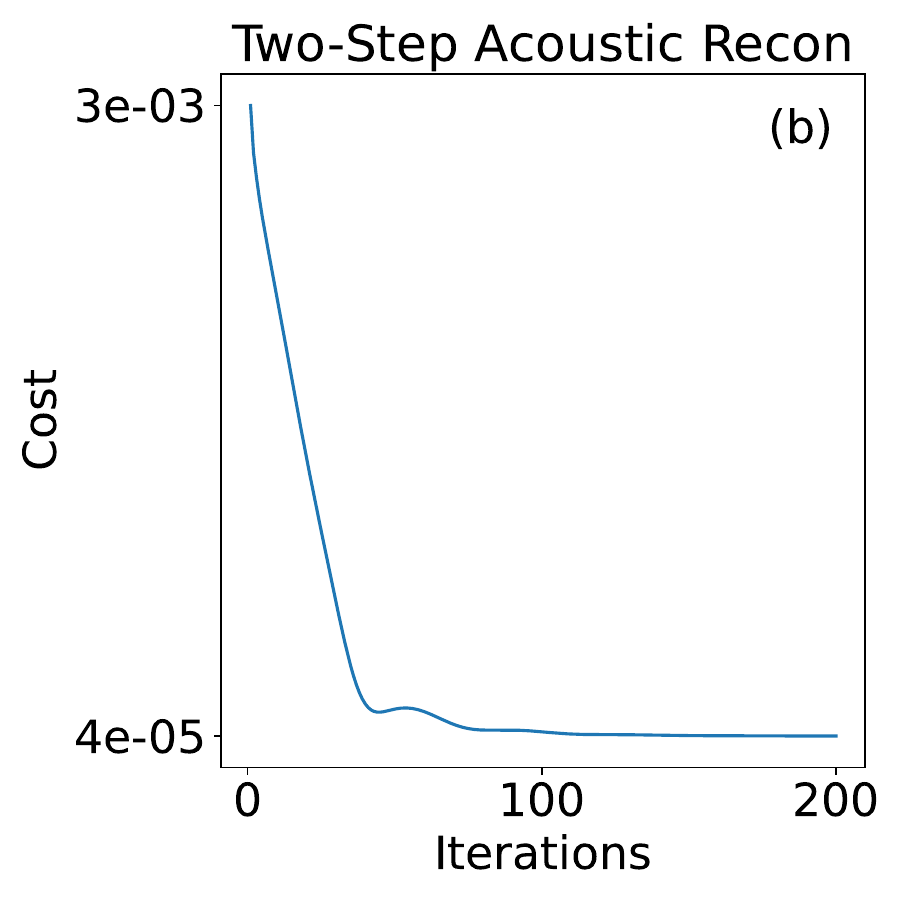}
    \includegraphics[width=0.32\linewidth]{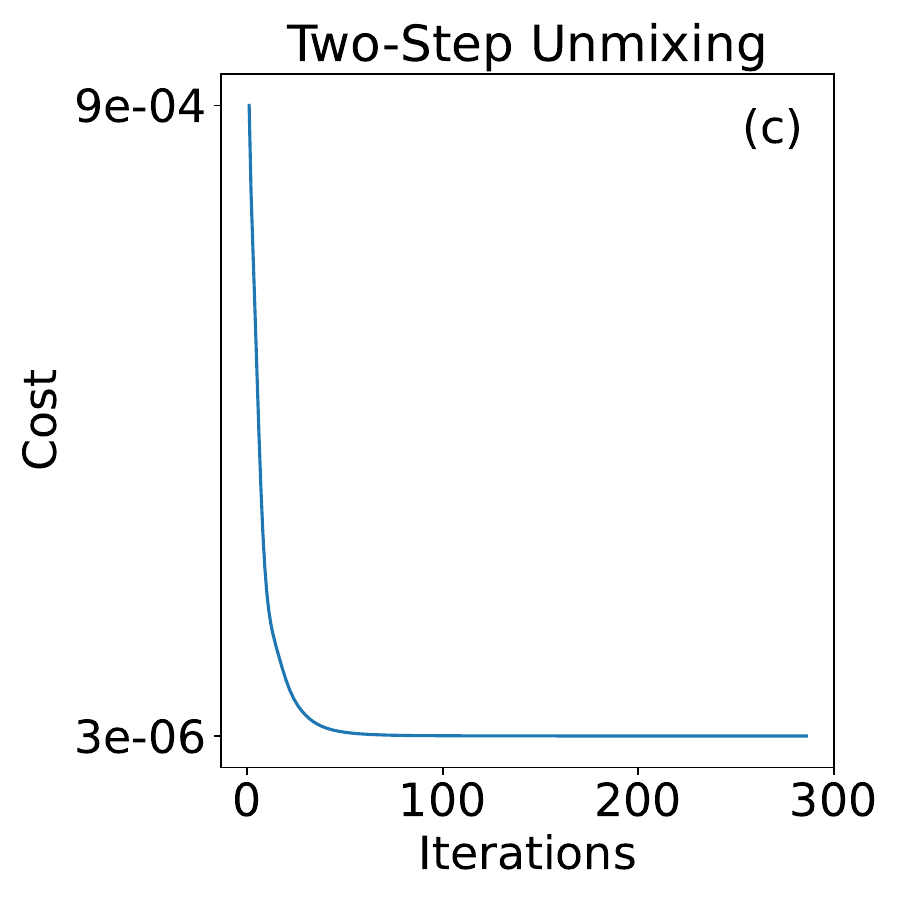}\\
    \includegraphics[width=0.32\linewidth]{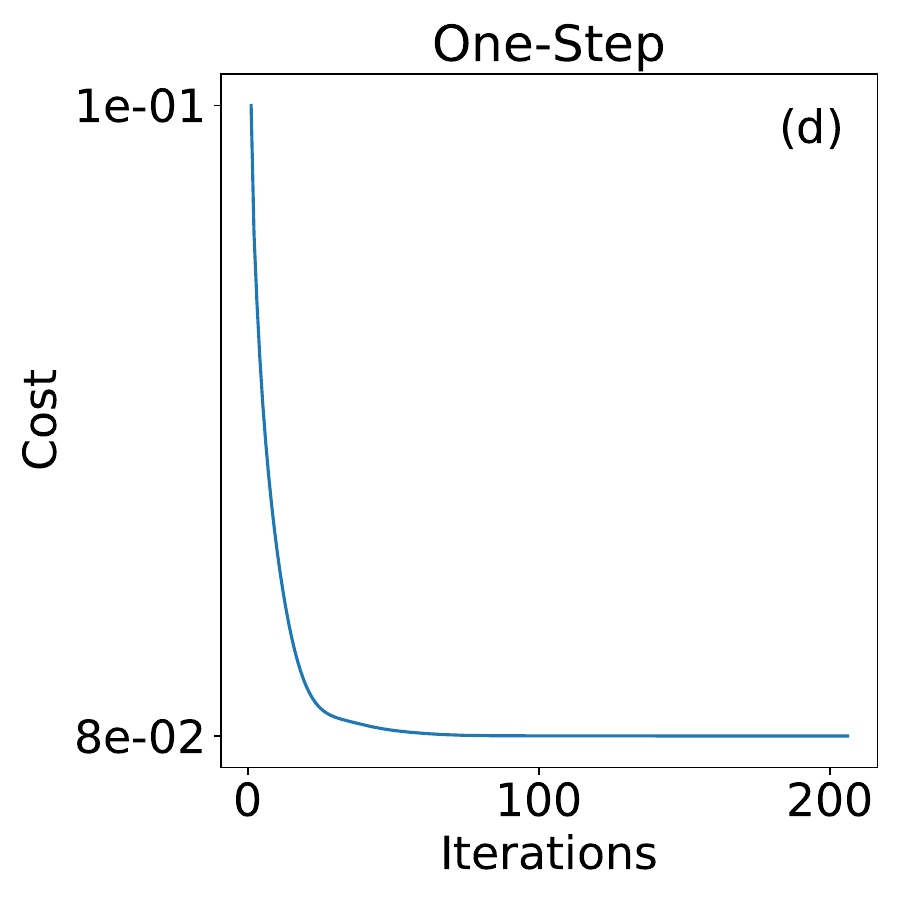}
    \includegraphics[width=0.32\linewidth]{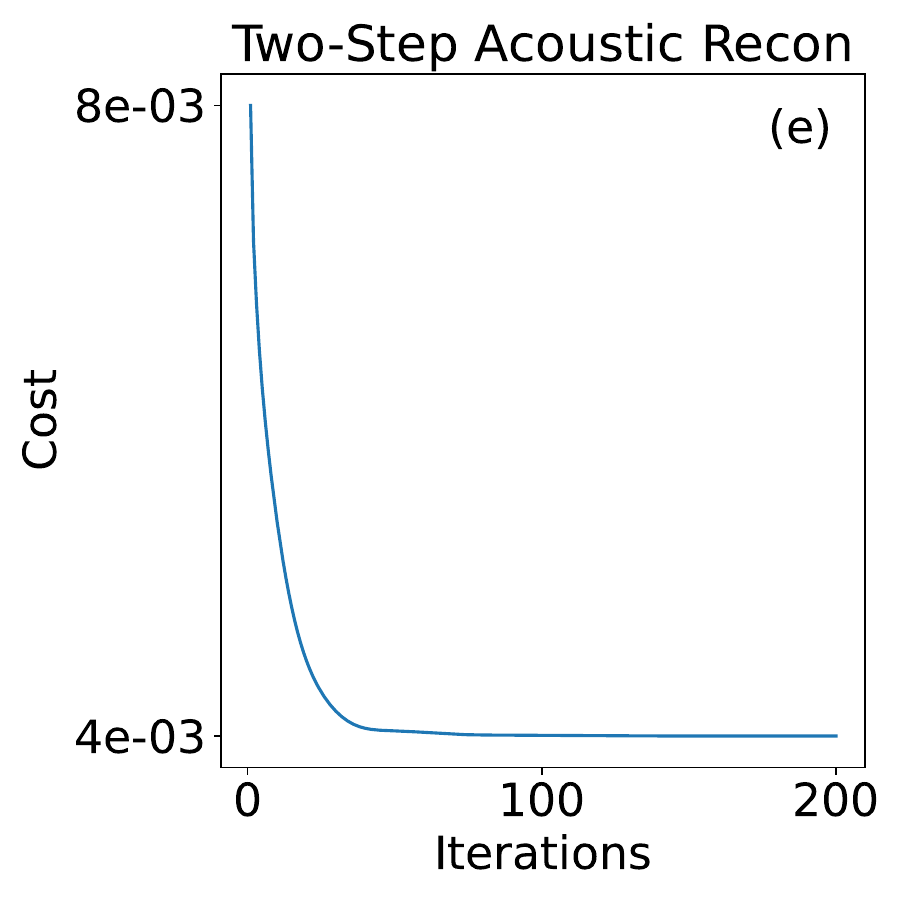}
    \includegraphics[width=0.32\linewidth]{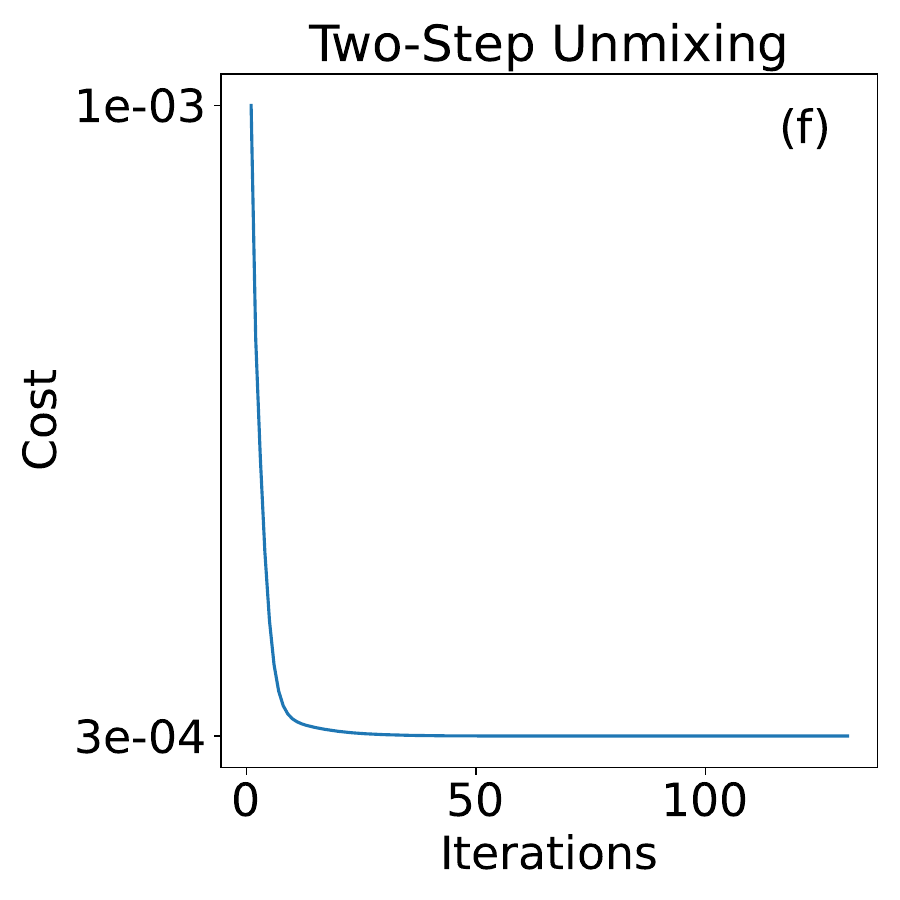}
    \caption{Cost evolution of the regularized approaches in the case of 1\% ((a)-(c)) and 10\% ((d)-(f)) noise in the measurement.}
    \label{fig:cost-evolution}
\end{figure}

This section contains additional figures and tables.
We show in Figure \ref{fig:cost-evolution} the cost evolution of the regularized one-step and two-step approaches shown in Section \ref{sec:low-noise} and \ref{sec:high-noise}.
In Table \ref{tab:photophysics}, we provide the physical properties of the experiment.

\begin{table}[h]
    \centering
    \begin{tabular}{c|ccc}
    \toprule
        Name 
        & $k$
        & $\varepsilon^{\text{ON}}$  & $\varepsilon^{\text{OFF}}$ \\
        \midrule
        Unit & $\text{s}^{-1}$ & $\mu\text{M}^{-1}\text{mm}^{-1}$ & $\mu\text{M}^{-1}\text{mm}^{-1}$ \\
        \midrule
        Species A &  4.0 & $5\times 10 ^{-3}$ & $1 \times 10^{-4}$ \\

        Species B &  2.0 & $4 \times 10^{-3}$ & $1 \times 10^{-4}$ \\

        Background  & 0 & $5 \times 10^{-3}$ & $5 \times 10^{-3}$ \\
        \bottomrule
    \end{tabular}
    \caption{Photo-physical properties of the two photo-switching reporter species and the background used in the simulations. Unit $\mu$M stands for micro-molar.}
    \label{tab:photophysics}
\end{table}

\section*{Appendix D}
Regarding the search for the hyperparameters,
the unregularized methods only have one hyperparameter, the number of iterations.
Because the linear systems $\mathbf{W}$, $\mathbf{S}$ and $\mathbf{A}$ are ill-conditioned (the estimated condition numbers $\kappa$ for the setups in Section \ref{sec:low-noise} and \ref{sec:high-noise} are $\kappa(\mathbf{W}) = 160$, $\kappa(\mathbf{S}) = 280$, and $\kappa(\mathbf{A}) = 460$),
we apply early stopping to regularize the solution and to avoid fitting it to noise.
The main hyperparameter of the regularized methods is the regularization weight.
They are tuned such that the reconstruction achieves the overall best PSNR for both species.
We used $K_2=20$ iterations for the inner loop of the proximal gradient step, threshold values of $\varepsilon_1=10^{-8}$ and $\varepsilon_2=10^{-4}$ for the stopping criteria, and a maximal number of iterations $K_1=300$ for the main problem in Algorithm \ref{alg:prox-outer}.

\section*{Appendix E}\label{sec:complexity}
We compare the complexity the two row operators on a single row at depth $m_2$ in the output image on CPU.
We assume that the acoustic image $q$ for one acoustic process and the kernel $h$ are squares of size $(M \times M)$ and $(K \times K)$, respectively.
The FFT approach uses 2D FFTs to compute correlations in the Fourier domain as
\begin{equation}
    \text{Extract}_{k_2} \left(\text{IFFT}_{\text{2D}}\left\{
    \text{FFT}_{\text{2D}}\left\{ q \right\}
    \cdot
    \text{FFT}_{\text{2D}}\left\{\text{conj}(h)\right\}
    \right\}\right),
\end{equation}
where $M$ and $K$ are of the same order, and conj() represents complex conjugate.
Its complexity is
\begin{equation}
    O\left(M^2\log M^2 + K^2\log K^2 + M^2\log M^2 \right) \sim O(M^2\log M).
\end{equation}
The custom method calculates correlations in the original domain at each pixel location $(m_1, m_2)$
\begin{equation}
    q[m_1, m_2] = \sum_{k_1=0}^{K-1} \sum_{k_2=0}^{K-1} \beta[k_1, k_2]  \eta[k_1-m_1, k_2-m_2].
\end{equation}
Its complexity is $O(K^2)$ per pixel and thus, $O(K^2 M)$ for a single row at depth $m_2$.

Further, our sparse representation of the kernel reduces the complexity for each pixel to $O(K)$ since the kernel is essentially a 1D curve (sparse representation) instead of a 2D image (dense representation).
Consequently, we have
\begin{equation}
    q[m_1, m_2] = \sum_{(k_1, k_2)\in \Lambda} \beta[k_1, k_2] \eta[k_1-m_1, k_2-m_2],
\end{equation}
where $\Lambda$ is a set of indices that approximate either the upper or lower semicircle.
The final complexity is thus $O(K M) \sim O(M^2)$ instead of $O(K^2 M) \sim O(M^3)$.
Then, we compare the complete acoustic spatial operator based on these two methods on CPU.
Their theoretical complexity is $O(M^3\log M)$ and $O(M^3)$, due to the for loop along the depth of size $M$.

\end{document}